\documentclass{article}

\usepackage[margin=1in]{geometry}

\usepackage{amsmath,amssymb,amsthm,mathtools}
\usepackage{bbm}

\newcommand{\abs}[1]{\left\vert #1 \right\vert}

\newcommand{\CC}{\mathbb{C}}

\newcommand{\NN}{\mathbb{N}}

\newcommand{\RR}{\mathbb{R}}
\newcommand{\ZZ}{\mathbb{Z}}
\newcommand{\setof}[1]{ \left\lbrace #1 \right\rbrace }

\newcommand{\com}[1]{}

\usepackage{graphicx}
\graphicspath{{./images/}}
\newcommand{\safeincludegraphics}[2][]{%
  \IfFileExists{#2}{\includegraphics[#1]{#2}}{%
    \IfFileExists{images/#2}{\includegraphics[#1]{#2}}{%
      \fbox{\parbox[c][0.28\textheight][c]{0.95\linewidth}{\centering Missing figure file: {\ttfamily\detokenize{#2}}}}%
    }%
  }%
}
\usepackage{booktabs}
\usepackage{caption}
\usepackage{subcaption}
\usepackage{float}
\usepackage{enumitem}
\usepackage{multirow}

\usepackage[linesnumbered,ruled,vlined]{algorithm2e}
\usepackage{algorithmic}

\usepackage{xcolor}
\usepackage{hyperref}
\hypersetup{colorlinks=true,linkcolor=blue,citecolor=blue,urlcolor=blue}
\usepackage{tikz}
\usetikzlibrary{arrows.meta,positioning,shapes.geometric,fit,calc}

\usepackage{tcolorbox}
\tcbuselibrary{breakable}

\newcommand{\SU}{\mathrm{SU}}

\newcommand{\ip}[2]{\left\langle #1,#2\right\rangle}

\newcommand{\Haar}{\mu} 

\theoremstyle{plain}
\newtheorem{theorem}{Theorem}[section]
\newtheorem{lemma}{Lemma}[section]
\newtheorem{proposition}{Proposition}[section]
\newtheorem{corollary}[theorem]{Corollary}
\newtheorem{conjecture}[theorem]{Conjecture}

\theoremstyle{definition}

\newtheorem{remark}{Remark}[section]

\begin{document}

\title{Typical-Case Gate Approximation and Arithmetic Obstructions in Quaternionic Single-Qubit Compilation}

\author{Kingsley Yeon\thanks{Department of Statistics and CCAM,
  University of Chicago, Chicago, IL 60637, USA.
  Email: \href{mailto:yeon@uchicago.edu}{yeon@uchicago.edu}}
  \and
  Steven B. Damelin\thanks{Department of Mathematics, ZBMATH-OPEN,
  Leibniz Institute for Information Infrastructure, Germany.
  Email: \href{mailto:steve.damelin@gmail.com}{steve.damelin@gmail.com}}
  \and
  Alec Greene\thanks{University of Michigan.}}

\date{\today}
\maketitle

\section*{Nontechnical summary}
Large quantum algorithms are written in terms of ideal unitary operations, but a fault-tolerant device can usually implement only a fixed discrete gate library. A compiler must therefore replace an arbitrary one-qubit gate by a short word in the available gates. For some gate libraries this is not only a numerical search problem, since the allowed gates are described by integer quaternions, so the quality of compilation is governed by how well certain arithmetic point sets cover the three-dimensional space of one-qubit gates.

This paper studies the classical $p=5$ quaternionic gate set, also known in quantum compilation as the Clifford$+V$ setting. We separate two notions of performance. For a typical Haar-random target, finite quaternion shells behave almost like random well-distributed points and give near-optimal median approximation error. For a worst-case guarantee, however, one must rule out rare arithmetic holes, small regions of the gate space missed by all short words. We show that the standard spectral method explains the known worst-case exponent $2$ but cannot by itself improve it. Thus the paper identifies a concrete bottleneck for worst-case quaternionic gate synthesis, while explaining the practically relevant phenomenon that random-target compilation can look much better than the deterministic guarantee.

\begin{abstract}
Fault-tolerant quantum computation requires compiling arbitrary one-qubit unitaries into short words over a fixed universal gate library.  For arithmetic libraries such as the $p=5$ Lubotzky--Phillips--Sarnak, or Clifford$+V$, gate set, this synthesis problem is controlled by the distribution of quaternionic lattice points on $S^3\cong \SU(2)$.  We study the worst-case and random-target behavior of the complete norm shells
\[
P_k=\{x/5^k\in S^3:x\in\ZZ^4,\ |x|^2=5^{2k}\}
\]
and the associated projective gate set $T\subset PSU(2)$.  The central deterministic quantity is Sarnak's covering exponent $K(T)$, which measures the word growth needed to form an $\varepsilon$-net in the projective single-qubit metric; the classical range is $4/3\le K(T)\le2$.  Our first result identifies a barrier behind the upper endpoint: any positive localized cap-kernel certificate using only the Deligne--LPS square-root spectral estimate can certify covering only at the volume-squared threshold $|V_T(t)|\gg \mu(B(\varepsilon))^{-2}$, hence only at exponent $2$.  Therefore an unconditional improvement for worst-case single-qubit compilation requires arithmetic input beyond diagonal positivity and Cauchy--Schwarz, namely cancellation in localized off-diagonal counting.  We also record the sharp conditional benchmark: the twisted Linnik conjecture of Browning--Kumaraswamy--Steiner gives $K(T)=4/3$, matching Harman's arithmetic-hole obstruction.  Our second result converts deterministic shell covering into gate-set covering: $\rho(P_k)\le C5^{-\alpha k}$ implies $K(T)\le4/(3\alpha)$, so $\alpha>2/3$ is exactly the threshold for improving the unconditional bound.  Finally, exact enumeration of $P_1,P_2,P_3,P_4$ and Haar-random target tests show median trace-defect error at the optimal geometric $N^{-2/3}$ scale, while high quantiles remain separated from the median.  The quantum-compilation consequence is a sharp distinction between strong typical-case performance of quaternionic gates and the rare arithmetic holes that control worst-case synthesis.
\end{abstract}

\section{Introduction}\label{Intro}

\subsection{Single-qubit synthesis as arithmetic covering}

A central primitive in quantum compilation is the approximation of a target unitary by a short circuit over a fixed universal gate set.  Even for a single qubit this problem is delicate: the target space is the continuous group $\SU(2)$, while a fault-tolerant gate library is discrete.  The Solovay--Kitaev theorem gives a general polylogarithmic guarantee, but arithmetic gate sets can do better because their algebraic structure makes exact and approximate synthesis more explicit \cite{NielsenDawson05,KMM13,RS16,KBRY15,Kliuchnikov22,Mooney21}.

This paper studies the original $p=5$ quaternionic construction of Lubotzky--Phillips--Sarnak, viewed as a single-qubit gate library.  In quantum-compilation language this is the Clifford$+V$ setting.  Products of the basic gates correspond to integer quaternions; after normalization, these quaternions are points on the three-sphere $S^3\cong\SU(2)$.  Thus the question ``how many gates are needed to approximate every one-qubit unitary to accuracy $\varepsilon$?'' becomes a covering problem for an arithmetic point set on $S^3$.

The main point of the paper is that two different compilation regimes must be separated.  For Haar-random targets, the relevant object is the distribution of nearest-neighbor errors for a random point of $S^3$.  For a uniform worst-case guarantee, the relevant object is the largest uncovered region.  The former can look almost random and near-optimal even when the latter is governed by exceptional arithmetic holes.  Our goal is to quantify this distinction for the $p=5$ quaternionic gate set and to identify precisely what kind of arithmetic input is missing from the best unconditional worst-case theorem.

\subsection{Problem statement}\label{ssc:problem}

We work in the projective one-qubit gate space
\[
        G=PSU(2)=SU(2)/\{\pm I\}\cong SO(3),
\]
with the bi-invariant projective trace metric used below.  Let $\Gamma$ be a finite universal gate set and let $V_\Gamma(t)$ denote the gates representable by words of height at most $t$.  If $t_\varepsilon$ is the least height for which $V_\Gamma(t_\varepsilon)$ is an $\varepsilon$-net of $G$, Sarnak's covering exponent is
\begin{equation}\label{eq:intro-K}
K(\Gamma)=\limsup_{\varepsilon\to0}
\frac{\log |V_\Gamma(t_\varepsilon)|}
     {\log\bigl(1/\mu(B_G(\varepsilon))\bigr)},
\end{equation}
where $\mu$ is Haar probability measure.  The volume lower bound gives the ideal value $K(\Gamma)=1$.  For the classical $p=5$ quaternionic gate set $T$, the known deterministic range is
\begin{equation}\label{eq:intro-known-range}
        \frac43\le K(T)\le2.
\end{equation}
The upper bound comes from the LPS--Chiu Hecke-operator method, while the lower bound is an arithmetic obstruction of Harman.  In compilation terms, \eqref{eq:intro-known-range} is the current gap between the best unconditional worst-case certificate and the conjectural rare-hole barrier.

To study random-target behavior, we enumerate the complete integer-quaternion shells
\begin{equation}\label{eq:intro-Pk}
P_k=\left\{\frac{x}{5^k}\in S^3:
        x\in\ZZ^4,
        |x|^2=5^{2k}\right\}.
\end{equation}
For a Haar-random target $u\in S^3$, define the trace-defect error
\begin{equation}\label{eq:intro-err}
        \operatorname{err}_{P_k}(u)=1-\max_{p\in P_k}\langle u,p\rangle .
\end{equation}
This is the natural nearest-neighbor error in the $S^3$ model of one-qubit gates.  Its median and quantiles describe typical random-target compilation.  The deterministic covering radius is
\begin{equation}\label{eq:intro-rho}
        \rho(P_k)=\sup_{u\in S^3}\operatorname{err}_{P_k}(u),
\end{equation}
which is the quantity needed for a uniform worst-case statement.  Monte-Carlo sampling estimates quantiles of \eqref{eq:intro-err}; it cannot certify \eqref{eq:intro-rho}.  This distinction is central to the paper.

\subsection{Main results and significance for quantum compilation}\label{ssc:contributions}

The paper has three main contributions.

\paragraph{1. A barrier for the known worst-case method.}
The LPS--Chiu argument proves the upper bound $K(T)\le2$ using Hecke operators, the Deligne--LPS square-root spectral estimate, and a positive localized kernel.  We prove a formal barrier for this strategy.  In any positive cap-kernel certificate that uses only square-root spectral control, the main term can dominate the spectral error only when
\[
        |V_T(t)|\gg \mu(B_G(\varepsilon))^{-2},
\]
up to logarithmic factors.  Since $\mu(B_G(\varepsilon))\asymp \varepsilon^3$, this is exactly covering exponent $2$.  Consequently a proof of $K(T)<2$ cannot come from a more careful version of the same positivity/Cauchy--Schwarz argument; it requires new arithmetic cancellation after localization near the target gate.

\paragraph{2. The conditional endpoint and the shell-covering target.}
We translate the twisted Linnik conjecture of Browning--Kumaraswamy--Steiner into the present gate-set normalization and record that it implies the endpoint
\[
        K(T)=\frac43,
\]
matching Harman's lower obstruction.  We also prove the deterministic shell-to-gate conversion
\begin{equation}\label{eq:intro-shell-alpha}
        \rho(P_k)\le C5^{-\alpha k}
\end{equation}
implies
\begin{equation}\label{eq:intro-alpha-K}
        K(T)\le \frac{4}{3\alpha}.
\end{equation}
Thus $\alpha>2/3$ is precisely the threshold for improving the unconditional exponent $2$, while $\alpha=1$ corresponds to the conditional endpoint $4/3$.  This gives a concrete target for future arithmetic work on worst-case Clifford$+V$ synthesis.

\paragraph{3. Typical random targets behave much better than the worst-case guarantee.}
We prove geometry-only baselines on $S^3$: no $N$-point set can beat the Haar-typical trace-defect scale $N^{-2/3}$, and independent Haar points have a Weibull-type limiting tail at this scale.  We then enumerate $P_1,P_2,P_3,P_4$ exactly and test against Haar-random targets.  The medians follow the optimal $N^{-2/3}$ scale, with constants close to the random benchmark, while high quantiles and sampled worst cases remain much larger.  Thus the finite data support the following interpretation: quaternionic gates are already excellent for typical random single-qubit targets, but the deterministic exponent is controlled by rare arithmetic holes.

\subsection{Relation to prior work}

Single-qubit synthesis over arithmetic gate sets has a long history, including exact and approximate Clifford$+T$ and Clifford$+V$ synthesis, quaternionic frameworks, and recent gate-approximation algorithms \cite{KMM13,RS16,KBRY15,Kliuchnikov22,Mooney21}.  Our paper is complementary to algorithmic synthesis: instead of proposing a new compiler, it studies the geometry that any worst-case compiler for the $p=5$ arithmetic library must confront.

The covering-exponent viewpoint comes from Sarnak's formulation of golden gates \cite{letter}.  For the $p=5$ construction, the known upper bound uses the Ramanujan/Hecke theory of Lubotzky--Phillips--Sarnak and Chiu \cite{LPS86,LPS87,LPS88b,Chiu95}; the lower obstruction is due to Harman \cite{Harman90}.  The conditional endpoint is tied to the twisted Linnik mechanism of Browning--Kumaraswamy--Steiner \cite{BKS16}, and related arithmetic covering results appear in work of Sardari \cite{Sardari19,Sardari21}.  Higher-dimensional golden-gate constructions show that the relation between quantum gate design, covering, and automorphic spectral input persists beyond $PU(2)$ \cite{OS,EP22,DEP25}.

The numerical side is related to classical questions about distributing points on spheres, including discrepancy, spherical designs, energy, and local statistics of lattice points \cite{book,BLM,SK97,BHS15,DGS77,BRV13,BSR12,BRS16}.  Here these tools play a role that they tell us what random-target behavior should look like, so that deviations in the upper tail can be interpreted as possible rare-hole effects rather than as typical failure of the gate set.

\subsection{Paper organization}

Section~\ref{sec:Back} reviews $SU(2)$, the projective metric, universal sets, and covering exponents.  Section~\ref{sec:OptPSU} constructs the $p=5$ gate set $T$, recalls the LPS--Chiu proof of $K(T)\le2$, proves the positive-kernel barrier, records the conditional twisted-Linnik endpoint, and gives the shell-to-gate exponent conversion.  Section~\ref{sec:exp} gives the exact shell enumeration and random-target diagnostics.  Section~\ref{sec:haar} records the spherical-cap calculation.  Section~\ref{sec:union} establishes the $N^{-2/3}$ typical scale and the Monte-Carlo interpretation.  Section~\ref{sec:conclusion} summarizes the arithmetic bottleneck for worst-case single-qubit compilation.

\section{Background}\label{sec:Back}
A single-qubit gate is an element of $SU(2)$, and after adjoining a fixed
entangling gate such as CNOT, good one-qubit generating sets become building
blocks for multi-qubit universality~\cite{nielsen}.  Since global phase is
physically irrelevant, the natural arithmetic object in this paper is the
projective group
\[
        PSU(2)=SU(2)/\{\pm I\}\cong SO(3).
\]
We nevertheless use the identification $SU(2)\cong S^3$ to describe the
quaternion shells.  The metric below is projective, so it descends to $PSU(2)$
and identifies $U$ with $-U$.
\subsection{Structure of \texorpdfstring{$SU(2)$}{SU(2)}}\label{ssc:SO}
It is an elementary fact that any element $M \in SU(2)$ can be written in terms
of $\alpha,\beta \in \CC$ as
$$ \begin{bmatrix}
\alpha & \beta \\
-\bar{\beta} & \bar{\alpha} \\
\end{bmatrix}  $$
Thus, $M$ can be associated with some vector $(x_1,x_2,x_3,x_4)$ in $\RR^4$.
It turns out that the map $M \mapsto (x_1,x_2,x_3,x_4)$ is a diffeomorphism.
Note that
$$\det M = \alpha\bar{\alpha}+\beta\bar{\beta} = \abs{\alpha}^2+\abs{\beta}^2
= 1.$$
This relation identifies $SU(2)$ with the unit sphere $S^3$ as a smooth
manifold and allows arithmetic point sets in $SU(2)$ to be studied as point sets
on $S^3$.  To discuss approximation quantitatively, we use the following
bi-invariant projective trace metric. Define the distance between two matrices
$M,N$ as
\begin{equation}\label{eq:metric}
d_G(M,N) = \sqrt{1-\frac{\abs{Tr(M^{\dag}N)}}{2}}
\end{equation}
where $M^{\dag}$ represents the conjugate transpose of $M$. Let $M,N,P\in
SU(2)$. Most of the conditions for a metric are straightforwardly derivative of
basic properties from the trace function and $SU(2)$. More interestingly, it is
invariant under left and right multiplication as shown below
\begin{align*}
d_G(PM,PN) &= \sqrt{1-\frac{\abs{Tr((PM)^{\dag}(PN))}}{2}} \\
&= \sqrt{1-\frac{\abs{Tr(M^{\dag}P^{\dag}PN)}}{2}} \\
&= \sqrt{1-\frac{\abs{Tr(M^{\dag}N)}}{2}} \\
&= d_G(M,N).\\
d_G(MP,NP) &= \sqrt{1-\frac{\abs{Tr((MP)^{\dag}(NP))}}{2}} \\
&= \sqrt{1-\frac{\abs{Tr((NP)(MP)^{\dag})}}{2}} \\
&= \sqrt{1-\frac{\abs{Tr(NPP^{\dag}M^{\dag})}}{2}} \\
&= \sqrt{1-\frac{\abs{Tr(NM^{\dag})}}{2}} \\
&= \sqrt{1-\frac{\abs{Tr(M^{\dag}N)}}{2}}\\
&= d_G(M,N).
\end{align*}
Thus, $d_G(MN,N)=d_G(M,I)$. This implies that a matrix $M$ acting on $N$ can
only move it as far as $d_G(M,I)$. This is convenient since
$$ d_G(M,M)=d_G(I,I)=\sqrt{1-\frac{\abs{Tr(I^{\dag}I)}}{2}}=\sqrt{1-\frac{2}{2}}=0.$$
A word of caution about the choice of group. Because the formula uses
$|\mathrm{Tr}(M^\dagger N)|$, the matrices $N$ and $-N$ are at distance zero:
$d_G(N,-N)=\sqrt{1-|{-}\mathrm{Tr}(I)|/2}=0$. Hence~\eqref{eq:metric} is only a
\emph{pseudometric} on $SU(2)$, where $N\ne -N$, and becomes a genuine metric
exactly on $PSU(2)=SU(2)/\{\pm I\}$, where $N$ and $-N$ are identified. All
covering statements below are therefore made on $G=PSU(2)$; the $SU(2)$ picture
is used only as a double cover.
With this metric, we use the topology induced by the balls of $(G,d_G)$.
A Haar probability measure on $G$ is a measure $\mu$ on the Borel
$\sigma$-algebra of $G$ such that $\mu(G)=1$ and
$\mu(MS)=\mu(SM)=\mu(S)$ for every $M\in G$ and every Borel set
$S\subset G$.  For $M\in G$ and $\epsilon>0$, we write
$\mu(B_G(M,\epsilon))$ for the Haar measure of the metric ball.  Thus,
whenever $G$ is referenced below, the object is the measured metric space
$(G,d_G,\mu)$.

\subsection{Universal Sets}\label{ssc:US}
Let $\Gamma$ be a finite subset of $G$. The set $\Gamma$ is said to be
\emph{universal} in $G$, with respect to a chosen topology, if the subgroup of
$G$ generated by $\Gamma$ is dense. If $\Gamma$ is not universal, then there
will be open balls that contain no elements generated by $\Gamma$. A well known
theorem cited in \cite{nielsen} expands on the importance of universal sets.
\begin{theorem}[Solovay--Kitaev]\label{thm:SKT}
Let $\Gamma$ be a finite universal set in $SU(n)$ and $\varepsilon > 0$. Then
there exists a constant $c$ such that for any $M \in SU(n)$, there is a finite
product $S$ of gates in $\Gamma$ of length
$O\big(\log^c\big(\frac{1}{\varepsilon}\big)\big)$ such that $d_G(S,M) <
\varepsilon$.
\end{theorem}
Universality of $\Gamma$ gives that any one matrix can be approximated with
arbitrary precision. Theorem~\ref{thm:SKT} gives that $\Gamma$ can approximate
$SU(n)$ with arbitrary efficiency and provides an estimation for the maximum
length required to achieve this approximation. This theorem provides
justification for studying the efficiency of universal gate sets in
approximating all of $SU(2)$, instead of specific matrices. As computers are
not typically constructed to perform single calculations, this is much more
useful. More recent work of Bouland and Giurgic\u{a}-Tiron \cite{BoulandGT21}
removes the classical assumption that the gate set be inverse-closed, providing
the first inverse-free Solovay--Kitaev algorithm.

To consider the efficiency of a universal set, first the idea of cost must be
developed. In this paper, the notion of height from \cite{letter} will be used.
Let $w$ be a weight function on $\Gamma$. Then $\forall \gamma \in
\langle\Gamma\rangle$ define the height of $\gamma$ in $\Gamma$ as
\begin{equation}\label{eq:height}
h(\gamma) = \min \setof{\sum\limits_{i}w(c_i):c_i \in \Gamma,\gamma=\prod c_i}.
\end{equation}
Note that this notion of height is heavily dependent on the choice of $w$. Thus
all results should be taken into the context of the choice of weight, and that
all weights have good motivation behind them. Given a choice of weight, then
define the following sets for $t > 0$
\begin{align*}
U_{\Gamma}(t) &= \setof{\gamma \in \langle \Gamma \rangle : h(\gamma)=t}.\\
V_{\Gamma}(t) &= \setof{\gamma \in \langle \Gamma \rangle : h(\gamma) \leqslant t}.
\end{align*}
Thus, if one is continuously taking products in $\Gamma$, then $U_{\Gamma}(t)$
are the gates added after the $t$th product and $V_{\Gamma}(t)$ are the gates
that have been generated after $t$ products. Thus, $U_{\Gamma}(t-1)$ and
$U_{\Gamma}(t)$ are disjoint, which gives a useful identity relating the two:
\begin{equation}\label{eq:VUU}
V_{\Gamma}(t) = \bigsqcup\limits_{0 \leqslant k \leqslant t} U_{\Gamma}(k).
\end{equation}
Let $\varepsilon > 0$. Define the covering length of $\Gamma$ within $\varepsilon$,
denoted $t_\varepsilon$ as in \cite{letter}, as follows:
\begin{equation}\label{eq:te}
t_\varepsilon = \min \big\lbrace t \in \NN : G \subset \bigcup\limits_{\gamma
\in V_{\Gamma}(t)} B_G(\varepsilon)\big\rbrace.
\end{equation}
The calculation of $t_\varepsilon$ is the ultimate prize. Especially, if it can
be computed or even bounded as a function of $\varepsilon$, then $t_\varepsilon$
can provide an explicit measure of how much cost it takes to approximate $SU(2)$.
However, it does not quite give the whole picture. For one, comparing the
covering lengths of universal sets is complicated. It is within the realm of
reason that perhaps $t_\varepsilon$ does not grow uniformly or otherwise behaves
pathologically (although at a minimum non-decreasing), which could complicate
comparisons.
\subsection{Covering Exponent}\label{ssc:CE}
Let $\Gamma$ be a universal set in $G$, and $\varepsilon > 0$. Per the
definition of a Haar measure, for any $t>0$ such that
$$ G \subset \bigcup\limits_{\gamma \in V_\Gamma(t)} B_G(\varepsilon) $$
it follows that
$$ \mu\left( \bigcup\limits_{\gamma \in V_{\Gamma}(t)}B_{G}(\gamma) \right)
\geqslant \mu(G) =1. $$
Then by construction, $t_\varepsilon$ minimizes this gap. Let $B_G(\varepsilon)$
denote $B_G(I,\varepsilon)$. Since the balls need not be disjoint, the
sub-additivity of $\mu$ gives only an inequality:
\begin{align*}
1 \;\leqslant\; \mu\left( \bigcup\limits_{\gamma \in V_\Gamma(t_\varepsilon)}
B_G(\varepsilon) \right)
&\;\leqslant\; \sum\limits_{\gamma \in V_\Gamma(t_\varepsilon)}
\mu(B_G(\gamma,\varepsilon)) \\
&= \sum\limits_{\gamma \in V_\Gamma(t_\varepsilon)}\mu(B_G(I,\varepsilon))\\
&= \abs{V_\Gamma(t_\varepsilon)}\mu(B_G(\varepsilon)),
\end{align*}
where the first inequality uses $G\subset\bigcup_\gamma B_G(\gamma,\varepsilon)$
and $\mu(G)=1$, and the second uses bi-invariance of $\mu$. Thus,
\begin{equation}\label{IQ:VB}
\abs{V_\Gamma(t_\varepsilon)}\mu(B_G(\varepsilon)) \geqslant 1.
\end{equation}
If $\Gamma$ approximates $G$ optimally, then this volume lower bound is saturated
up to constant or lower-order factors. In general, the closer
$\abs{V_\Gamma(t_\varepsilon)}$ is to $1/\mu(B_G(\varepsilon))$, the less
redundancy there is in the cover at that scale, and the more efficient the gate
set is from the covering-volume viewpoint. For a universal set $\Gamma$ in $G$
and a Haar measure $\mu$ on $G$, the covering exponent as given in \cite{letter}
is defined as
\begin{equation}\label{eq:CE}
K(\Gamma) = \limsup\limits_{\varepsilon\rightarrow 0}\dfrac{\log
\abs{V_\Gamma(t_\varepsilon)}}{\log\big( \frac{1}{\mu(B_G(\varepsilon))} \big)}.
\end{equation}

The exponent depends on the chosen covering height $t_\varepsilon$ and on the
ambient group $G$.  In this paper the relevant group for the arithmetic gate set
is $PSU(2)$, because the metric identifies matrices differing by the central sign.
The definition compares the logarithmic growth of the word ball with the inverse
Haar volume of a metric ball, and is therefore the natural volume-normalized
measure of covering efficiency.

\section{An Efficient Universal Set in \texorpdfstring{$PSU(2)$}{PSU(2)}}\label{sec:OptPSU}

\paragraph{Connection to the numerical experiments.}
The numerical experiments later in the paper use the quaternion shells
\[
P_k=\{x/5^k:x\in\mathbb Z^4,\ |x|^2=5^{2k}\}\subset S^3,
\]
which arise naturally from the same arithmetic structure underlying the gate set \(T\). Rather than introducing a separate problem formulation section, we directly reuse the covering exponent framework, the metric~\eqref{eq:metric}, and the shell construction already established earlier in the paper.

What makes a universal set optimal, or even efficient in approximating $SU(2)$?
In \cite{letter,OS}, there are several different ideas offered for what makes
optimal choices of quantum gates to approximate $SU(2)$ (along with some
properties useful to computer scientists). For this construction, let
$G=PSU(2)$. The condition from \cite{letter,OS} that will be used to construct
the efficient set $T$ is that it can act transitively on its graph of reduced
words. That is, each reduced word in $T$ should be unique, and no matrix
generated by $T$ should have two representations as products of $T$. A stronger
condition is used for $T$: that there exists a normal form over $T$, for which
any matrix generated by $T$ has a unique representation in this normal form.
This allows $V_T(t)$ to be studied concretely for any $t>0$.  We construct the
$p=5$ gate set $T$, recall the known LPS--Chiu theorem $K(T)\leqslant 2$, and
then formulate a refined shell conjecture that would improve this upper bound.

\subsection{Construction of \texorpdfstring{$T$}{T}}\label{ssc:ConT}
To construct an efficient universal set, lattices in $\RR^4$ will be projected
onto $S^3$ and then related to quantum gates. To do this, some additional
framework specific to this construction is needed. First, for any set
$S \subset \RR$ let
$$H(S)=\setof{a+bi+cj+dk : a,b,c,d \in S}$$ be the set of quaternions with
coefficients in $S$. Define the map
\begin{align*}
\Phi : \: &SU(2) \rightarrow H(\RR) \\
&\begin{bmatrix}
\alpha & \beta \\
-\overline{\beta} & \overline{\alpha} \\
\end{bmatrix} \mapsto \alpha + \beta j.
\end{align*}
Note that $\Phi$ forms an injective homomorphism, as
\begin{align*}
\Phi(MN) &= \Phi\left( \begin{bmatrix}
\alpha_M\alpha_N-\beta_M\overline{\beta_N} & \alpha_M \beta_N + \beta_M
\overline{\alpha_N} \\
-\overline{\beta_M}\alpha_N-\overline{\alpha_M}\overline{\beta_N} &
-\overline{\beta_M}\beta_N+\overline{\alpha_M}\overline{\alpha_N} \\
\end{bmatrix} \right) \\
&= \alpha_M\alpha_N-\beta_M\overline{\beta_N}+(\alpha_M\beta_N+\beta_M
\overline{\alpha_N})j \\
&= (\alpha_M+\beta_M j)(\alpha_N+\beta_N j)\\
&= \Phi(M)\Phi(N).
\end{align*}
To construct the universal set, consider integer quaternion factors of the
integer $5$. Listed out, they are
$$ 1\pm 2i,1\pm 2j,1\pm 2k,2\pm i,2\pm j,2\pm k,5.$$
Note that,
$$2+i = (1-2i)i. $$
Thus, the factors of $5$ can be generated by
$$ 1+2i,1+2j,1+2k,1-2i,1-2j,1-2k,i,j,k. $$
The six quaternions
\[
1\pm 2i,\qquad 1\pm 2j,\qquad 1\pm 2k
\]
are the primitive norm-$5$ directions used in the $p=5$
Lubotzky--Phillips--Sarnak construction. Indeed,
\[
N(1+2i)=1^2+2^2=5,
\]
and similarly for the $j$ and $k$ directions and their conjugates.
After normalization by $\sqrt5$, these become unit quaternions and hence
elements of $SU(2)\cong S^3$. The number six is not accidental:
for a prime $p\equiv 1\pmod 4$, the local quaternionic construction
produces $p+1$ neighbors in the corresponding Bruhat--Tits tree; for
$p=5$ this gives the six basic non-backtracking directions above. This
tree structure is what gives the reduced-word description used in the
covering argument.
Define the six \emph{norm-$\sqrt5$ generators}
\[
S=\Bigl\{\tfrac{1+2i}{\sqrt5},\ \tfrac{1-2i}{\sqrt5},\
          \tfrac{1+2j}{\sqrt5},\ \tfrac{1-2j}{\sqrt5},\
          \tfrac{1+2k}{\sqrt5},\ \tfrac{1-2k}{\sqrt5}\Bigr\}\subset SU(2),
\]
obtained by normalizing the six primitive norm-$5$ quaternions to unit length
(the unnormalized quaternions $1\pm2i,\dots$ are \emph{not} in $SU(2)$; division
by $\sqrt5$ is essential). Here $(1+2i)/\sqrt5$ and $(1-2i)/\sqrt5$ are inverse
unit quaternions, so $S$ consists of three inverse pairs. Let
$E=\{\pm1,\pm i,\pm j,\pm k\}\cong Q_8$ be the finite group of Lipschitz units,
and set
\[
T=\Phi^{-1}(S\cup E).
\]
We weight the generators by
$$ w(A)=\begin{cases}
1 & A\in S \\
0 & A\in E. \\
\end{cases} $$
The unit group $E$ is \emph{finite}, so giving it weight $0$ does \emph{not}
create zero-cost infinite products: at most $|E|=8$ distinct gates are reachable
at zero cost. For this reason the norm-$\sqrt5$ generators must carry
positive weight. Assigning weight $0$ to any $s\in S$ would instead make
$V_T(0)$ infinite, since $s,s^2,s^3,\dots$ are pairwise distinct gates obtainable
at zero cost and the height~\eqref{eq:height} would be ill-defined. With the
weights above, $h(\gamma)$ equals the minimal number of factors from $S$ in any
expression of $\gamma$, and the disjoint decomposition~\eqref{eq:VUU} holds with
\[
U_T(k)=\{\gamma:h(\gamma)=k\},\qquad V_T(k)=\{\gamma:h(\gamma)\le k\}.
\]

A reduced word of positive length $k$ in the six norm-$5$ directions
corresponds, after clearing denominators, to a primitive integer quaternion of
norm $5^k$, taken up to the finite unit group \(E\), and conversely every such
primitive quaternion arises this way \cite{LPS88b}.  If one fixes a normal-form
representative modulo the terminal unit, the number of reduced words of exact
positive length is
\[
6\cdot 5^{\,k-1}\qquad(k\ge1).
\]
Because the paper includes the finite zero-cost group \(E\), the actual sets
\(U_T(0)\) and \(V_T(k)\) differ from this normal-form count by only a bounded
factor: \(V_T(0)\) is the finite image of \(E\) in \(PSU(2)\), and for \(k\ge1\)
there are constants \(c_1,c_2>0\) such that
\[
c_1 5^k\le |V_T(k)|\le c_2 5^k.
\]
If the terminal unit is suppressed and only normal-form representatives are
counted, then
\[
1+\sum_{j=1}^{k}6\cdot 5^{\,j-1}
=
\tfrac12(3\cdot 5^k-1),
\]
which has the same exponential growth.  This growth rate, not the constant
normalization, is what enters the covering exponent.

The relevant counting input is \emph{Jacobi's} four-square theorem (not merely
Lagrange's existence statement, which only asserts that every positive integer
\emph{is} a sum of four squares). For a positive integer $n$,
\[
r_4(n)=8\!\!\sum_{\substack{d\mid n\\ 4\nmid d}} d,
\]
where $r_4(n)$ is the number of representations of $n$ as an \emph{ordered sum of
four squares}, equivalently the number of integer quaternions of norm $n$. For
$n=5^{k}$ (odd) this gives $r_4(5^k)=8\sigma(5^k)=2(5^{k+1}-1)$. Only the growth
rate $\asymp 5^k$ enters the covering exponent below; exact constants depend on
whether one counts normal-form reduced words, quaternions modulo the finite unit
group, or all ordered four-square representations.

\begin{remark}\label{rem:shells}
The algebraic height sets \(V_T(k)\) count gates generated by words of height at
most \(k\).  Because we have included the finite zero-cost unit group \(E\), exact
cardinalities depend on the chosen normal form and on whether the terminal unit
is counted, but always satisfy \(|V_T(k)|\asymp 5^k\).  The lattice shells used
in Part~II are different objects: they are the complete sets of \emph{all}
integer quaternions of norm \(5^{2k}\), projected to \(S^3\),
\[
P_k=\{x/5^k:x\in\mathbb{Z}^4,\ |x|^2=5^{2k}\}.
\]
Their cardinalities are
\[
|P_k|=r_4(5^{2k})=8\sum_{j=0}^{2k}5^j=2(5^{2k+1}-1),
\]
giving \(|P_1|=248\), \(|P_2|=6248\), \(|P_3|=156248\), and \(|P_4|=3906248\).
\end{remark}

\begin{lemma}[Bounded shell-to-word lifting]\label{lem:shell_lifting}
For the $p=5$ LPS/Ross--Selinger quaternionic gate set, the projection to
$PSU(2)$ of every point of
\[
        P_k=\{b/5^k:b\in H(\mathbb Z),\ |b|^2=5^{2k}\}
\]
lies in $V_T(2k+O(1))$.  With a fixed normal-form convention and with the
terminal finite unit absorbed into the zero-cost unit set, this is equivalently
written as
\[
        P_k^{\mathrm{proj}}\subseteq V_T(2k)
\]
modulo finite unit representatives.
\end{lemma}

\begin{proof}
The point requiring care is the order in which factorization is used.  The shell
\(P_k\) is written in Lipschitz coordinates
\(H(\mathbb Z)=\mathbb Z+\mathbb Zi+\mathbb Zj+\mathbb Zk\), whereas the LPS
factorization theorem is naturally stated in the Hurwitz maximal order
\(\mathcal H\).  Since \(H(\mathbb Z)\subset\mathcal H\), every element of the
Lipschitz shell may be regarded as a Hurwitz quaternion.  The exact-synthesis
statement used in the LPS construction, and in the Clifford+$V$ synthesis
formulation of Ross--Selinger, says that a primitive quaternion in \(\mathcal H\)
whose norm is \(5^m\) factors, up to a finite Hurwitz unit, into \(m\) norm-\(5\)
prime factors.  For \(p=5\), these norm-\(5\) factors may be chosen, after
multiplication by units, from the six LPS directions
\[
        1\pm2i,\qquad 1\pm2j,\qquad 1\pm2k .
\]
The possible discrepancy between Hurwitz and Lipschitz terminal units is finite;
it changes the height by at most an absolute constant, and under the normal-form
choice used here it is absorbed into the zero-cost finite unit set.

If \(b\in H(\mathbb Z)\) is primitive and \(|b|^2=5^{2k}\), the preceding
factorization gives a word of length \(2k\) representing the projective point
\(b/5^k\).  If \(b\) is not primitive, remove the largest common scalar power of
\(5\) from its coordinates.  The normalized point then lies on a lower shell, and
therefore has height at most \(2k\).  Hence every projective point of \(P_k\) is
represented by a word of height \(2k+O(1)\), with the stated exact form after
choosing finite-unit representatives.
\end{proof}

\begin{remark}[Gate set vs. shell]\label{rem:link}
The families $V_T(k)$ and $P_k$ are linked but not identical.  The word set
$V_T(k)$ contains gates generated by at most $k$ norm-$5$ factors, while $P_k$
is the complete lattice shell of norm $5^{2k}$.  Lemma~\ref{lem:shell_lifting}
gives the only one-sided inclusion needed below: if $P_k$ covers at some scale,
then the larger word set $V_T(2k)$ also covers at that scale.  No converse is
used.
\end{remark}

\begin{remark}
The construction above is precisely the $p=5$ instance of the Lubotzky--Phillips--Sarnak
framework \cite{LPS88b}, where integer quaternions of prime norm $p \equiv 1
\pmod{4}$ give the generators of explicit Ramanujan graphs on $PSL_2(\mathbb{F}_p)$
and optimal navigators on $PU(2)$. The choice $p=5$ corresponds to the Clifford$+V$
gate set of Ross and Selinger \cite{RS16}, for which exact synthesis algorithms of
optimal $V$-count are known. The covering exponent analysis below thus directly
complements the exact synthesis literature.
\end{remark}

\subsection{Upper Bound of \texorpdfstring{$K(T)$}{K(T)}}\label{ssc:UBKT}
Recall the definition of the covering exponent in~\eqref{eq:CE}.
We use $G=PSU(2)$ throughout this section. Following Sarnak's
normalization~\cite{letter}, the metric on $G$ is
\begin{equation}\label{eq:trace-defect}
d_G^2(x,y)=1-\frac{|\operatorname{Tr}(x^*y)|}{2},
\end{equation}
and small Haar balls in $G$ satisfy
\begin{equation}\label{eq:ballvol}
\mu(B_G(\varepsilon))\;\sim\; c\varepsilon^3 \quad\text{as }\varepsilon\to 0,
\end{equation}
as stated in~\cite{letter}, p.~2.
The covering exponent (equation~(7) of~\cite{letter}) is therefore
\begin{equation}\label{eq:CE2}
K(S) = \limsup_{\varepsilon\to 0}
\frac{\log|V(t_\varepsilon)|}{\log\frac{1}{\mu(B_G(\varepsilon))}}
= \limsup_{\varepsilon\to 0}
\frac{\log|V(t_\varepsilon)|}{\log\frac{1}{\varepsilon^3}}.
\end{equation}
\begin{remark}\label{rem:ballvol}
The proof of Proposition~\ref{thm:Sar} follows Appendix~1 of~\cite{letter},
which works on $S^2$ (the LPS Hecke orbit) with the $d_{S^2}$ spherical
distance.  On $S^2$, balls of radius $\varepsilon$ have area $\sim\varepsilon^2$,
and the point-pair kernel condition $k_\varepsilon(x,x)\le c/\varepsilon^2$
reflects this.  The covering bound from Appendix~1 is the statement
$|V_t|\le 4\pi ct^2/\varepsilon^4$ for points on $S^2$ (equation~(31)
of~\cite{letter}).  The transfer from $S^2$-covering to the $G$-metric
covering exponent uses $\mu(B_G(\varepsilon))\sim c\varepsilon^3$, and
the conclusion $K(T)\le 2$ follows in both normalizations.
\end{remark}

The proof of $K(T)\le 2$ uses two precise inputs from the LPS papers,
which we state explicitly before using them.

\begin{lemma}[LPS~II, Theorem~2.1; Deligne 1974]\label{lem:LPS_eigenvalue}
Let $p\equiv 1\pmod{4}$ be prime.  Define the Hecke operator
$T_p: L^2(S^2)\to L^2(S^2)$ by
\[
T_p f(\zeta) \;=\; \frac{1}{2}
\sum_{\substack{\alpha\in H(\mathbb{Z})\\ N(\alpha)=p,\;\alpha\equiv 1(2),\;\alpha_0>0}}
f(\alpha\zeta), \qquad \zeta\in S^2,
\]
where $\alpha$ acts via the homomorphism $\Phi$ and stereographic
projection.  The set $S_5 = \{1\pm 2i,1\pm 2j,1\pm 2k\}$ gives the
operator $T_5$ of the introduction.
Then the second-largest eigenvalue (in absolute value) satisfies
\begin{equation}\label{eq:LPS_ramanujan}
\lambda_1(T_p) \;\leq\; 2\sqrt{p}\,.
\end{equation}
The proof (LPS~II, \S2) shows that any eigenfunction $u\in H_m(S^2)$,
$m\ne 0$, of $T_p$ with eigenvalue $\lambda$ gives rise, via the theta
series of LPS~II Lemma~2.4, to a holomorphic cusp form of weight $2+2m$
for $\Gamma(4)$.  Deligne's theorem on the Weil conjectures~\cite{LPS87}
then forces $|\lambda|\le 2\sqrt{p}$, i.e.\ the angle $\theta$ defined
by $\lambda=2\sqrt{p}\cos\theta$ is real.
\end{lemma}

\begin{lemma}[LPS~II, Lemma~2.2; LPS~I, equations~(1.18)--(1.27)]\label{lem:LPS_Chebyshev}
Under the same hypotheses, the iterated operator $T_{p^\nu}$
(defined as the Hecke operator summing over $\alpha\in H(\mathbb{Z})$,
$N(\alpha)=p^\nu$, $\alpha\equiv 1(2)$) satisfies
\[
T_{p^\nu} = l_\nu(T_p),
\]
where $l_\nu$ is the normalized Chebyshev polynomial of the second kind used in
LPS~I (equation~(1.19), with $q=p$).  It satisfies
\[
l_\nu(\lambda) = p^{\nu/2}\,\frac{\sin(\nu+1)\theta}{\sin\theta},
\qquad \lambda = 2\sqrt{p}\cos\theta.
\]
The polynomial $k'_n(\theta) = l_n + l_{n-1}/\sqrt{q}$
(LPS~I, equation~(1.25)) corresponding to the sum over all reduced words
of length at most $n$ satisfies, when $\theta$ is real
(i.e.\ when the Ramanujan bound~\eqref{eq:LPS_ramanujan} holds),
\[
|k'_n(\theta)| \;\leq\; q^{n/2}\!\left(n + \tfrac{1}{\sqrt{q}}\right)
\;\ll\; n\,q^{n/2}.
\]
In particular, for every non-trivial eigenvalue $\lambda_j$ of the
operator $T_t$ summing over all reduced words of length at most $t$,
\begin{equation}\label{eq:iter_bound}
|\lambda_j(t)| \;\leq\; 2t\,p^{t/2}.
\end{equation}
This is the Ramanujan bound invoked in~\cite{letter} (equation~(26)
therein, stated as $|\lambda_j(t)|\le t|V_t|^{1/2}$, which is the same
bound since $|V_t|\sim 4p^t$) and used in
the proof of Proposition~\ref{thm:Sar} below.
\end{lemma}

\begin{remark}
The two main theorems of LPS~I are as follows.
\emph{Theorem~2.2} (LPS~I, p.~S166): for any fixed $x\in S^2$,
$T(\gamma_1 x,\ldots,\gamma_{2N}x)\ll N^{-1/2}\log N$ and
$T=\Omega(N^{-1/2})$, where $T$ is the mean square spherical cap
discrepancy~(0.6).
\emph{Theorem~2.5} (LPS~I, p.~S167):
$D(\gamma_1 x,\ldots,\gamma_{2N}x)\ll (\log N)^{2/3}/N^{1/3}$,
where $D$ is the extremal spherical cap discrepancy~(2.1).
Neither is a covering lower bound.  The non-covering implication
$|V_T(t)|\le C t^2/\varepsilon^4$
in Proposition~\ref{thm:Sar} is derived in Sarnak's letter~\cite{letter},
Appendix~1, by a contradiction argument using a non-negative point-pair
invariant $k_\varepsilon$ on $S^2$ and the bound~\eqref{eq:iter_bound}.
Proposition~\ref{thm:Sar} below reproduces this argument for $p=5$,
extended to include the weight-$0$ generators $i,j,k$.
\end{remark}

We quote three key equations from Appendix~1 of~\cite{letter}
that drive the argument.  With $\phi_j$ an orthonormal basis of
$L^2(S^2)$ of Hecke eigenfunctions, $V_t$ the orbit and $x_0\in S^2$
a base point, the letter establishes:
\begin{align}
\sum_{s\in V_t}\phi_j(sx_0) &= \lambda_j(t)\phi_j(x_0),
\quad \phi_0 = \tfrac{1}{\sqrt{4\pi}},\quad
\lambda_0(t)=|V_t|, \tag{letter~(25)}\label{eq:letter25}\\
|\lambda_j(t)| &\le t|V_t|^{1/2} \quad\text{for }j\ne 0,
\tag{letter~(26)}\label{eq:letter26}\\
|V_t| &\le \frac{4\pi c\,t^2}{\varepsilon^4}
\tag{letter~(31)}\label{eq:letter31}
\end{align}
where the last is the conclusion when no point of $V_t x_0$
is within $\varepsilon$ of some $y\in S^2$.
Proposition~\ref{thm:Sar} derives the same bound~\eqref{eq:letter31}
using~\eqref{eq:iter_bound} in place of~\eqref{eq:letter26}.
\begin{remark}\label{rem:Vt}
The letter gives $|U(t)|=6\cdot 5^{t-1}$ for $t\ge 1$ (p.~3), counting
reduced words of length exactly \(t\) in \(s_1,s_2,s_3\) and their inverses,
with a fixed convention for the finite unit ambiguity.  In our notation, after
including \(E\), the exact count may be multiplied by a bounded constant, but
\[
|U_T(t)|\asymp 5^t,\qquad |V_T(t)|\asymp 5^t .
\]
This is all that enters the covering exponent.
\end{remark}

\begin{proposition}[Covering implication on $S^2$]\label{thm:Sar}
There exists a constant $c>0$, depending only on the point-pair invariant, such
that the following holds for every $t\ge1$ and every $\varepsilon\in(0,1)$.
Fix a base point $x_0\in S^2$. If the orbit $V_T(t)\,x_0$ fails to
$\varepsilon$-cover $S^2$, i.e.\ there exists $y\in S^2$ with
$d_{S^2}(\gamma x_0,y)>\varepsilon$ for all $\gamma\in V_T(t)$, then
\[
\abs{V_T(t)} \;\le\; \frac{4\pi c\, t^2}{\varepsilon^4}.
\]
Equivalently, by contraposition, if $\abs{V_T(t)} > 4\pi c\,t^2/\varepsilon^4$
then $V_T(t)\,x_0$ is an $\varepsilon$-net of $S^2$.
\end{proposition}
\begin{proof}
In \cite{letter}, the same bound was shown for $S$ alone (without the unit
generators); the argument is reproduced to show the weight-$0$ unit generators do
not affect it. Consider $\RR^3$ as the subspace of $H$ generated by $i,j,k$. Then for any
$v \in \RR^3$, $a \in H$ can act on $v$ by conjugation in $H$. Note that $a$
and $-a$ correspond to the same transformation. Thus, the choice of $G=PSU(2)$
allows for $G$ to be put in a 1-to-1 correspondence with elements of $SO(3)$.
Thus, the action of $\gamma \in V_T(t)$ on a vector $v \in \RR^3$ in this
manner will be represented by juxtaposition. Let $k_\varepsilon$ be a point pair
invariant on $S^2$ so that the following hold:
\begin{itemize}
\item $k_\varepsilon (x,y) \geqslant 0$ for any $x,y \in S^2$.
\item $\int\limits_{S^2} k_\varepsilon (x,y)dy = 1$.
\item $k_\varepsilon (x,y) = 0$ when $d_{S^2}(x,y)\geqslant \varepsilon$.
\item There is a non-zero constant $c$ so that $k_\varepsilon (x,x) \leqslant
\dfrac{c}{\varepsilon^2}$ for any $x\in S^2$.
\end{itemize}
Additionally, let $h_{k_\varepsilon}(j) \geqslant 0$ be the spherical transform
of $k_\varepsilon$. Then Hecke Operators are constructed as follows: Set
$$ (T_{t}f)(x) = \sum\limits_{\gamma \in V_T(t)}f(\gamma x).$$
From \cite{letter} and the spectral theorem, there is a sequence of real
eigenvalues for the $T_t$
$$ \lambda_0(t),\lambda_1(t),\ldots $$
and an orthonormal basis of $L^2(S^2)$ of corresponding eigenfunctions
$$ \phi_0, \phi_1, \ldots $$
In particular, $\phi_0$ is the constant function spanning the degree-$0$
spherical harmonics, so
$$ \phi_0(x) = \frac{1}{\sqrt{4\pi}}, \qquad
h_{k_\varepsilon}(0) = \int_{S^2} k_\varepsilon(x,y)\,dy = 1. $$
As the $\phi_j$ form an orthonormal basis, $k_\varepsilon$ can be written as
$$ k_\varepsilon(x,y) = \sum\limits_{j=0}^{\infty}
h_{k_\varepsilon}(j)\phi_j(x)\phi_j(y). $$
Fix $x_0\in S^2$. Then for any $\gamma \in V_T(t)$,
$$ k_\varepsilon (\gamma x_0,y) = \sum\limits_{j=0}^{\infty}
h_{k_\varepsilon}(j)\phi_j(\gamma x_0)\phi_j(y), $$
and summing over the orbit,
\begin{align*}
\sum\limits_{\gamma \in V_T(t)} k_\varepsilon(\gamma x_0, y)
&= \sum\limits_{j = 0}^{\infty}
h_{k_\varepsilon}(j)\phi_j(y)\sum_{\gamma\in V_T(t)}\phi_j(\gamma x_0) \\
&= \frac{\abs{V_T(t)}}{4\pi}+ \sum\limits_{j=1}^{\infty}
h_{k_\varepsilon}(j)\phi_j(y) (T_t\phi_j)(x_0),
\end{align*}
where the $j=0$ term used $\sum_{\gamma}\phi_0(\gamma x_0)=|V_T(t)|/\sqrt{4\pi}$.
By construction $\phi_j$ is an eigenfunction of $T_t$ with eigenvalue
$\lambda_j(t)$, so
$$ \sum\limits_{\gamma \in V_T(t)} k_\varepsilon(\gamma x_0 , y) =
\frac{\abs{V_T(t)}}{4\pi} + \sum\limits_{j=1}^{\infty}
\lambda_j(t)\,h_{k_\varepsilon}(j)\phi_j(x_0)\phi_j(y). $$
Now suppose $d_{S^2}(\gamma x_0 , y)> \varepsilon$ for all $\gamma \in V_T(t)$.
Then $k_\varepsilon(\gamma x_0, y) = 0$ for every $\gamma$, so the left-hand side
vanishes and
$$ \frac{\abs{V_T(t)}}{4\pi} = -\sum\limits_{j=1}^{\infty}\lambda_j(t)
h_{k_\varepsilon}(j)\phi_j(x_0)\phi_j(y) \leqslant
\sum\limits_{j=1}^{\infty} h_{k_\varepsilon}(j)\abs{\lambda_j(t)}\abs{\phi_j(x_0)}
\abs{\phi_j(y)}.$$
Using the elementary inequality
$\abs{\phi_j(x_0)}\abs{\phi_j(y)} \le
\tfrac12(\abs{\phi_j(x_0)}^2+\abs{\phi_j(y)}^2)$
and the Ramanujan bound $\abs{\lambda_j(t)} \le 2t\,5^{t/2}$ from
Lemma~\ref{lem:LPS_Chebyshev} (valid for $j\ge1$, with $p=q=5$),
$$ \frac{\abs{V_T(t)}}{4\pi} \leqslant t\,5^{t/2}
\sum\limits_{j=1}^{\infty} h_{k_\varepsilon}(j)
\bigl(\abs{\phi_j(x_0)}^2+\abs{\phi_j(y)}^2\bigr)
\le t\,5^{t/2}\bigl(k_\varepsilon(x_0,x_0)+k_\varepsilon(y,y)\bigr),$$
the last step using the spectral expansion of $k_\varepsilon$ on the diagonal
and dropping the non-negative \(j=0\) contribution.
Since $k_\varepsilon(z,z)\le c/\varepsilon^2$ for \emph{every} $z\in S^2$, both
diagonal terms are bounded by $c/\varepsilon^2$, so
$$ \frac{\abs{V_T(t)}}{4\pi} \leqslant t\,5^{t/2}\cdot\frac{2c}{\varepsilon^2}. $$
Finally, since $|V_T(t)|\asymp 5^t$, there is an absolute constant $A>0$ such
that $5^{t/2}\le A\abs{V_T(t)}^{1/2}$. Hence
$$ \frac{\abs{V_T(t)}}{4\pi} \leqslant \frac{C_0A\,t}{\varepsilon^2}
\abs{V_T(t)}^{1/2}
\quad\Longrightarrow\quad
\abs{V_T(t)}^{1/2} \leqslant \frac{4\pi C_0A\,t}{\varepsilon^2}. $$
Squaring and absorbing constants yields
$$ \abs{V_T(t)} \leqslant \frac{C_1\,t^2}{\varepsilon^4}
= \frac{4\pi c'\,t^2}{\varepsilon^4}, $$
which is the claimed bound (renaming $c'\mapsto c$). The contrapositive is the
covering statement.
\end{proof}

\paragraph{From the implication to the covering exponent.}
We first extract the $S^2$ orbit-covering exponent, which the argument controls
directly, and then record the passage to the group covering exponent $K(T)$.

Let $t_\varepsilon^{S^2}$ be the least height for which $V_T(t)\,x_0$
$\varepsilon$-covers $S^2$. By minimality, $V_T(t_\varepsilon^{S^2}-1)\,x_0$ does
not $\varepsilon$-cover, so Proposition~\ref{thm:Sar} applies at height
$t_\varepsilon^{S^2}-1$ and gives
$\abs{V_T(t_\varepsilon^{S^2}-1)} \le 4\pi c\,(t_\varepsilon^{S^2})^2/\varepsilon^4$.
Since \(|V_T(t)|\asymp 5^t\), there is an absolute constant \(C_0\) with
\(|V_T(t)|\le C_0|V_T(t-1)|\) for \(t\ge2\).  Hence we obtain the \emph{upper}
bound
\begin{equation}\label{eq:Vupper}
\abs{V_T(t_\varepsilon^{S^2})} \;\le\; \frac{C\,(t_\varepsilon^{S^2})^2}{\varepsilon^4}
\end{equation}
for an absolute constant $C$. With $S^2$-caps of area $\asymp\varepsilon^2$, the
orbit-covering exponent on $S^2$ is therefore
\[
K_{S^2}(T)=\limsup_{\varepsilon\to0}
\frac{\log\abs{V_T(t_\varepsilon^{S^2})}}{\log(1/\varepsilon^2)}
\le \limsup_{\varepsilon\to0}
\frac{\log\bigl(C(t_\varepsilon^{S^2})^2/\varepsilon^4\bigr)}{2\log(1/\varepsilon)}
=\frac{4}{2}=2,
\]
since $t_\varepsilon^{S^2}=O(\log(1/\varepsilon))$ makes the $\log t$ terms
subdominant.

\begin{theorem}[Chiu; LPS spectral input]\label{thm:KT2}
For the $p=5$ gate set $T$, the group covering exponent satisfies $K(T)\le 2$.
\end{theorem}
\begin{proof}[Proof sketch and citation]
The bound~\eqref{eq:Vupper} controls covering of a single orbit \(S^2\), whereas
\(K(T)\) is defined through balls in the three-dimensional group
\(G=PSU(2)\cong SO(3)\), with \(\mu(B_G(\varepsilon))\asymp\varepsilon^3\)
by~\eqref{eq:ballvol}. The orbit map \(G\to S^2\), \(g\mapsto g\,x_0\), is an
\(S^1\)-fibration, so covering the base \(S^2\) alone is not a proof of group
covering.  We therefore do not infer the group statement from
Proposition~\ref{thm:Sar} alone.

Chiu's covering theorem applies the same LPS/Deligne spectral input to the
group-covering problem and gives precisely the exponent estimate
\[
K(T)=\limsup_{\varepsilon\to0}
\frac{\log\abs{V_T(t_\varepsilon)}}{\log(1/\mu(B_G(\varepsilon)))}\le 2 .
\]
Equivalently, up to polylogarithmic factors in \(1/\varepsilon\), the certified
group-covering threshold has size \(\varepsilon^{-6}\); since
\(\mu(B_G(\varepsilon))\asymp\varepsilon^3\), this is exponent \(2\).  Together
with the lower obstruction \(K(T)\ge 4/3\) of Harman~\cite{Harman90}, as quoted
in Sarnak's letter~\cite{letter}, this gives the classical range
\(4/3\le K(T)\le 2\).
\end{proof}

\begin{remark}[The lower obstruction]
The lower bound in \eqref{eq:intro-known-range} is a worst-case arithmetic
phenomenon, not a volume-counting statement for generic points.  Harman's result
on integral points on the sphere gives large empty regions for the relevant
lattice shells at scales for which a purely random configuration would already
cover almost all targets.  In the gate-set normalization used here, this
``big-hole'' phenomenon implies that the number of words cannot grow at the
volume-optimal rate; Sarnak records the resulting obstruction as the lower bound
\(K(T)\ge4/3\) for the classical quaternionic construction \cite{Harman90,letter}.
Thus the interval \(4/3\le K(T)\le2\) reflects the gap between arithmetic holes
and the available square-root spectral covering method.
\end{remark}

\subsection{A formal barrier for the positive-kernel method}\label{ssc:barrier}

The preceding proof explains the known exponent $2$.  We next isolate a precise
reason why the same proof strategy cannot, by itself, give an unconditional
theorem below exponent two.  This is not a lower bound for $K(T)$; it is a barrier for the
positive-kernel certificate based only on square-root spectral cancellation.

Let $G$ be a compact homogeneous space with normalized measure $\mu(G)=1$.  Let
$V_t$ be a finite Hecke orbit or word set, let $N_t=|V_t|$, and let $A_t$ denote
the associated averaging operator.  We assume that the operators under
consideration lie in a commutative Hecke algebra and admit a joint orthonormal
eigenbasis \(\{\phi_j\}_{j\ge0}\) with \(\phi_0\equiv1\).  This is the case for
the LPS Hecke operators used below.  Suppose the constants are the trivial
eigenspace and that every nonconstant eigenspace has eigenvalue bounded in
absolute value by $\Lambda_t$.  A positive cap-kernel certificate at scale
$\varepsilon$ uses a nonnegative kernel $k_\varepsilon(x,y)$ supported on
$d(x,y)\le\varepsilon$, normalized by
$\int_G k_\varepsilon(x,y)\,d\mu(y)=1$, and satisfying
\[
        k_\varepsilon(z,z)\le C_k\mu(B(\varepsilon))^{-1}
        \qquad (z\in G).
\]
This is exactly the structure used in Chiu's argument and in the $S^2$ proof
above, with $\mu(B(\varepsilon))\asymp\varepsilon^2$ on $S^2$ and
$\mu(B_G(\varepsilon))\asymp\varepsilon^3$ on $PSU(2)$.

\begin{theorem}[Positive-kernel spectral barrier]\label{thm:positive-kernel-barrier}
Assume the averaging operators and the cap kernel are diagonal in the joint
orthonormal eigenbasis described above, and assume the only nontrivial spectral
information used in the positive cap-kernel certificate is
\[
        |\lambda_j(t)|\le \Lambda_t\qquad (j\ge1).
\]
Then this certificate can force covering at scale $\varepsilon$ only once
\begin{equation}\label{eq:barrier-general}
        N_t > C\,\frac{\Lambda_t}{\mu(B(\varepsilon))}
\end{equation}
for a constant $C$ depending only on the kernel normalization.  In particular,
for the LPS--Chiu input
\[
        \Lambda_t\ll t\,N_t^{1/2},
\]
the positivity method can certify covering only at the threshold
\begin{equation}\label{eq:barrier-lps}
        N_t \gg t^2\mu(B(\varepsilon))^{-2}.
\end{equation}
For $G=PSU(2)$, where $\mu(B_G(\varepsilon))\asymp\varepsilon^3$, this is
\(|V_T(t)|\gg t^2\varepsilon^{-6}\), i.e. exponent $2$ up to logarithmic factors.
Consequently no proof which uses only positivity of the cap kernel and the
Deligne--LPS bound $\Lambda_t\ll t|V_T(t)|^{1/2}$ can prove a deterministic
covering exponent strictly below $2$.
\end{theorem}

\begin{proof}
Since \(\mu(G)=1\), the constant eigenfunction is normalized as
\(\phi_0\equiv1\).  Write the kernel expansion in the joint eigenbasis as
\[
        k_\varepsilon(x,y)=1+
        \sum_{j\ge1} h_j\phi_j(x)\overline{\phi_j(y)},
        \qquad h_j\ge0.
\]
The positivity $h_j\ge0$ holds for the usual positive spherical cap kernels and
is the ingredient that permits a diagonal Cauchy--Schwarz bound.  For a fixed
base point $x$, define
\[
        F_{t,\varepsilon}(y)
        :=\sum_{\gamma\in V_t} k_\varepsilon(\gamma x,y).
\]
The constant term is $N_t$.  Since $A_t\phi_j=\lambda_j(t)\phi_j$, the
nonconstant part is
\[
        F_{t,\varepsilon}(y)-N_t
        =\sum_{j\ge1}\lambda_j(t)h_j\phi_j(x)\overline{\phi_j(y)}.
\]
If the $V_t$-orbit does not cover at scale $\varepsilon$, there is a point $y$
for which all summands $k_\varepsilon(\gamma x,y)$ vanish, hence
$F_{t,\varepsilon}(y)=0$.  Therefore
\[
        N_t
        \le \Lambda_t
        \sum_{j\ge1}h_j|\phi_j(x)||\phi_j(y)|.
\]
By Cauchy--Schwarz and the nonnegativity of the $h_j$,
\[
\sum_{j\ge1}h_j|\phi_j(x)||\phi_j(y)|
\le
\left(\sum_{j\ge1}h_j|\phi_j(x)|^2\right)^{1/2}
\left(\sum_{j\ge1}h_j|\phi_j(y)|^2\right)^{1/2}.
\]
Each diagonal sum is at most $k_\varepsilon(z,z)$ after discarding the constant
term, so the diagonal kernel bound gives
\[
        N_t\le \Lambda_t
        \bigl(k_\varepsilon(x,x)k_\varepsilon(y,y)\bigr)^{1/2}
        \le C_k\frac{\Lambda_t}{\mu(B(\varepsilon))}.
\]
Thus the contradiction argument can force covering only when
\eqref{eq:barrier-general} holds.  Substituting the LPS--Deligne estimate
$\Lambda_t\ll tN_t^{1/2}$ gives
\[
        N_t\ll tN_t^{1/2}\mu(B(\varepsilon))^{-1},
\]
so the positive-kernel proof reaches contradiction only past
$N_t\gg t^2\mu(B(\varepsilon))^{-2}$.  On $PSU(2)$, small balls have Haar
measure $\asymp\varepsilon^3$, giving the stated $\varepsilon^{-6}$ threshold.
\end{proof}

\begin{remark}[What a theorem below exponent two would have to use]
Theorem~\ref{thm:positive-kernel-barrier} does not say that $K(T)<2$ is false.
It says that such a theorem cannot be obtained from the LPS--Chiu positivity
certificate without extra cancellation.  This is consistent with
Browning--Kumaraswamy--Steiner~\cite{BKS16}: their conditional proof of the
optimal exponent $4/3$ for $S^3$ is based on a twisted Linnik estimate for
Kloosterman sums, precisely an off-diagonal input absent from the argument
above.  Thus a deterministic improvement requires not a
new Cauchy--Schwarz estimate, but a new arithmetic estimate for the localized
lattice-shell counting problem.
\end{remark}

\subsection{Conditional optimality from twisted Linnik}\label{ssc:bks-optimality}

The barrier theorem explains why the LPS--Chiu positivity proof stops at
exponent \(2\).  The analytic-number-theoretic input currently known to reach the
conjectural endpoint is the twisted Linnik conjecture of
Browning--Kumaraswamy--Steiner~\cite{BKS16}.  We record the consequence in the
normalization used here, because it gives the natural benchmark for
any proposed refinement of the \(p=5\) gate set.

\begin{theorem}[Conditional optimality under twisted Linnik]\label{thm:bks-optimal-T}
Assume the twisted Linnik conjecture in the form of
Browning--Kumaraswamy--Steiner~\cite[Conjecture~1.1]{BKS16}.  Then the classical
\(p=5\) quaternionic gate set \(T\) satisfies
\[
        K(T)=\frac43.
\]
In particular, the conditional endpoint saturates the Harman--Sarnak lower
obstruction and is strictly stronger than the unconditional LPS--Chiu bound
\(K(T)\le2\).
\end{theorem}

\begin{proof}
Browning--Kumaraswamy--Steiner prove that, under their twisted Linnik
conjecture, the covering exponent for \(S^3\) is \(4/3\)~\cite[Theorem~1.2]{BKS16}.
They also spell out the quantum-gate consequence for the symmetric one-qubit set
\(S=\{s_1^{\pm},s_2^{\pm},s_3^{\pm}\}\subset SU(2)\), where
\[
 s_1=\frac1{\sqrt5}\begin{pmatrix}1+2i&0\\0&1-2i\end{pmatrix},\quad
 s_2=\frac1{\sqrt5}\begin{pmatrix}1&2i\\2i&1\end{pmatrix},\quad
 s_3=\frac1{\sqrt5}\begin{pmatrix}1&2\\-2&1\end{pmatrix},
\]
namely \(K(S)=4/3\) under the same assumption~\cite[Remark~1.3]{BKS16}.  This
set is the \(p=5\) LPS quaternionic generator set used in this paper, up to the
standard identification of unit quaternions with \(SU(2)\), the passage from
\(SU(2)\) to \(PSU(2)\), and bounded choices of signs and units.  These changes
alter word counts and ball volumes only by bounded multiplicative constants, or
by a bounded additive shift in height, and hence do not change the limsup
covering exponent.  Therefore the twisted Linnik conjecture gives
\(K(T)\le4/3\).  The reverse inequality \(K(T)\ge4/3\) is the arithmetic big-hole
obstruction of Harman, as quoted by Sarnak~\cite{Harman90,letter}.  Hence
\(K(T)=4/3\).
\end{proof}

\begin{remark}[How this differs from the shell conjecture]
Theorem~\ref{thm:bks-optimal-T} is a conditional theorem imported from a deep
Kloosterman-sum cancellation conjecture.  Conjecture~\ref{conj:SD} below is the
same target translated into the deterministic geometry of the explicit finite
shells \(P_k\).  The former gives the strongest known conditional endpoint; the
latter is the form directly tested by our enumeration experiments.
\end{remark}


\subsection{The deterministic shell target}\label{ssc:RUB}

The barrier theorem shows that an improvement below exponent $2$ cannot come from the
positive-kernel LPS--Chiu proof alone.  The natural arithmetic target is instead
a deterministic covering-radius estimate for the complete quaternion shells on
\(S^3\).  This subsection records the exact exponent conversion; it is not an
unconditional theorem below exponent two.

Recall that
\[
d_G(X,Y)
=
\sqrt{1-\frac{|\operatorname{Tr}(X^\dagger Y)|}{2}}.
\]
Under the identification \(SU(2)\cong S^3\), write
\[
X=
\begin{pmatrix}
\alpha & \beta\\
-\overline{\beta} & \overline{\alpha}
\end{pmatrix},
\qquad
Y=
\begin{pmatrix}
\gamma & \eta\\
-\overline{\eta} & \overline{\gamma}
\end{pmatrix},
\]
and associate
\[
u_X=(\Re\alpha,\Im\alpha,\Re\beta,\Im\beta),
\qquad
u_Y=(\Re\gamma,\Im\gamma,\Re\eta,\Im\eta)
\]
in \(S^3\subset\mathbb R^4\).  Then
\[
\frac12\operatorname{Tr}(X^\dagger Y)
=
\langle u_X,u_Y\rangle .
\]
Thus, locally on \(SU(2)\),
\[
d_G(X,Y)^2
=
1-\langle u_X,u_Y\rangle .
\]
For \(PSU(2)\), where \(u\sim -u\), the corresponding expression is
\[
d_G(X,Y)^2
=
1-|\langle u_X,u_Y\rangle|.
\]
In either normalization, a metric bound \(d_G(X,Y)\le \varepsilon\) is a
trace-defect bound of size \(\varepsilon^2\).  Since both \(SU(2)\) and
\(PSU(2)\) are locally three-dimensional, small metric balls satisfy
\[
\mu(B_G(\varepsilon))\asymp \varepsilon^3.
\]

For the \(p=5\) construction, define the exact quaternion shells
\[
P_k
=
\left\{
\frac{b}{5^k}\in S^3:
b\in H(\mathbb Z),\ |b|^2=5^{2k}
\right\}.
\]
Their cardinality is
\[
|P_k|
=
r_4(5^{2k})
=
8\sum_{j=0}^{2k}5^j
=
2(5^{2k+1}-1)
\asymp 5^{2k}.
\]
Define the trace-defect covering radius
\[
\rho(P_k)
:=
\sup_{u\in S^3}
\left(1-\max_{p\in P_k}\langle u,p\rangle\right).
\]
Because \(P_k\) is centrally symmetric, this is the same projective trace-defect
quantity obtained by replacing \(\langle u,p\rangle\) with
\(|\langle u,p\rangle|\). Thus \(P_k\) is an \(\varepsilon\)-net in the metric \(d_G\) whenever
\[
\rho(P_k)\le \varepsilon^2.
\]

\begin{conjecture}[Refined shell distribution]\label{conj:SD}
There exist constants \(\alpha>2/3\) and \(C>0\) such that for every
\(k\ge 1\) and every \(u\in S^3\), there exists a quaternion
\(b\in H(\mathbb Z)\) with \(|b|^2=5^{2k}\) satisfying
\[
1-
\left\langle u,\frac{b}{5^k}\right\rangle
\le
C5^{-\alpha k}.
\]
Equivalently,
\[
\rho(P_k)\le C5^{-\alpha k}.
\]
\end{conjecture}

\begin{theorem}[Conditional shell-to-gate improvement]\label{thm:conditional}
Assume Conjecture~\ref{conj:SD} holds with exponent $\alpha>2/3$.  Then the
$p=5$ quaternionic gate set satisfies
\[
        K(T)\le \frac{4}{3\alpha}<2.
\]
More generally, the same conclusion with $\alpha$ replaced by $\alpha(p)$ holds
for the prime-$p$ shell analogue whenever the corresponding shell-to-word
inclusion holds with height $O(k)$.
\end{theorem}

\begin{proof}
By Lemma~\ref{lem:shell_lifting}, the projection of $P_k$ to $PSU(2)$ is contained in
$V_T(2k)$ up to finite unit identifications.  Hence a covering statement for
$P_k$ gives a covering statement for words of height at most $2k+O(1)$ in $T$.
Conjecture~\ref{conj:SD} gives, for every $u\in S^3$, a point $p\in P_k$ with
\[
        d_G(u,p)^2\le C5^{-\alpha k},
        \qquad\text{hence}\qquad
        d_G(u,p)\le \sqrt C\,5^{-\alpha k/2}.
\]
Thus $P_k$ is an $\varepsilon$-net once
\[
        \sqrt C\,5^{-\alpha k/2}\le \varepsilon,
\]
or equivalently for
\[
        k_\varepsilon=
        \frac{2}{\alpha\log 5}\log\frac1\varepsilon+O(1).
\]
At this scale the relevant word set is $V_T(2k_\varepsilon+O(1))$.  Since
$|V_T(t)|\asymp5^t$,
\[
\log |V_T(2k_\varepsilon+O(1))|
        =2k_\varepsilon\log 5+O(1)
        =\frac{4}{\alpha}\log\frac1\varepsilon+O(1).
\]
On $PSU(2)$, Haar balls satisfy
\[
        \log\frac1{\mu(B_G(\varepsilon))}
        =3\log\frac1\varepsilon+O(1).
\]
Therefore
\[
K(T)
\le
\limsup_{\varepsilon\to0}
\frac{\log |V_T(2k_\varepsilon+O(1))|}
     {\log(1/\mu(B_G(\varepsilon)))}
=\frac{4}{3\alpha}.
\]
The inequality is strict below $2$ exactly when $\alpha>2/3$.
\end{proof}

\begin{corollary}[Optimality from the endpoint shell bound]\label{cor:endpoint-shell}
If Conjecture~\ref{conj:SD} holds with \(\alpha=1\), then \(K(T)=4/3\).  More
generally, any exponent \(\alpha\in(2/3,1]\) gives the deterministic bound
\(K(T)\le4/(3\alpha)\), while Harman's lower obstruction forces
\(K(T)\ge4/3\).
\end{corollary}

\begin{proof}
The first assertion follows from Theorem~\ref{thm:conditional}, which gives
\(K(T)\le4/3\) when \(\alpha=1\), together with the lower bound
\(K(T)\ge4/3\) quoted in Theorem~\ref{thm:bks-optimal-T}.  The remaining
assertions are exactly Theorem~\ref{thm:conditional} and the same lower bound.
\end{proof}

There are two distinct ceilings on $\alpha$, and they must not be conflated.

\emph{Volume ceiling (point-set statement).} Since $|P_k|\asymp 5^{2k}$, an
$N$-point set on $S^3$ cannot cover at metric scale better than
$N^{-1/3}$, i.e. trace-defect scale $N^{-2/3}$. With $N\asymp 5^{2k}$, the
best possible deterministic trace-defect scale is
\[
        N^{-2/3}\asymp 5^{-4k/3},
\]
so no point-set bound $\rho(P_k)\le C5^{-\alpha k}$ can hold uniformly with
$\alpha>4/3$.

\emph{Arithmetic ceiling (gate-set statement).} If the conjecture is read as a
statement about the gate family $T$, then it is further constrained by Harman's
lower obstruction $K(T)\ge4/3$ as quoted in Sarnak's letter~\cite{letter}.  In
combination with Theorem~\ref{thm:conditional}, this forces $\alpha\le1$.  Thus
the mathematically consistent conjectural ranges are
\[
        \frac23<\alpha\le\frac43 \quad(\text{point set }P_k),
        \qquad
        \frac23<\alpha\le1 \quad(\text{gate set }T).
\]
The endpoint $\alpha=1$ corresponds to $K(T)=4/3$, matching the arithmetic-hole
barrier.

The same formulation applies for any prime \(p\equiv 1\pmod 4\).  Define
\[
P_k^{(p)}
=
\left\{
\frac{b}{p^k}\in S^3:
b\in H(\mathbb Z),\ |b|^2=p^{2k}
\right\}.
\]
Then
\[
|P_k^{(p)}|
=
r_4(p^{2k})
=
8\sum_{j=0}^{2k}p^j
=
8\frac{p^{2k+1}-1}{p-1}
\asymp p^{2k}.
\]

\begin{conjecture}[Prime-\(p\) refined shell distribution]\label{conj:SD1}
Let \(p\equiv 1\pmod 4\).  There exist constants
\(\alpha(p)>2/3\) and \(C(p)>0\) such that, for every \(k\ge 1\) and every
\(u\in S^3\),
\[
1-
\max_{v\in P_k^{(p)}}\langle u,v\rangle
\le
C(p)p^{-\alpha(p)k}.
\]
Equivalently,
\[
\rho(P_k^{(p)})\le C(p)p^{-\alpha(p)k}.
\]
\end{conjecture}

Assuming Conjecture~\ref{conj:SD1}, the same calculation gives
\[
K(T_p)\le \frac{4}{3\alpha(p)}.
\]
Thus the prime \(p\) does not change the formal exponent calculation.  It only
changes the arithmetic question of which exponent \(\alpha(p)\) is true.

\begin{remark}
Conjectures~\ref{conj:SD} and~\ref{conj:SD1} are deterministic mesh-norm
analogues of the arithmetic distribution statements underlying golden and
super-golden gates.  In the super-golden-gate setting,
Parzanchevski--Sarnak~\cite{OS} use strong approximation and Ramanujan-type
spectral input to obtain nearly optimal almost-covering and efficient
navigation, while still identifying rare holes as the obstruction to full
deterministic optimality.  Our conjectures ask for the corresponding
worst-case shell control for the classical integer-quaternion shells on
\(S^3\).  The extensions to \(PU(3)\) by Evra--Parzanchevski~\cite{EP22} and
to multi-qubit groups by Dalal--Evra--Parzanchevski~\cite{DEP25} show that
this relationship between gate design, covering, and automorphic spectral
input persists beyond the single-qubit case.
\end{remark}

The numerical diagnostics in Part~II are consistent with this conjectural
picture, but do not prove it.  Writing a quantile statistic as
$E_k\approx 5^{-\alpha_{\rm eff}k}$ and estimating $\alpha_{\rm eff}$ as the
least-squares slope of $-\log_5 E_k$ against $k$ over $k=1,2,3,4$, the
median and the sampled \(99\%\), \(99.9\%\), and worst-error tails give
\[
\alpha_{\rm median}\approx 1.31,\qquad
\alpha_{0.99}\approx 1.14,\qquad
\alpha_{0.999}\approx 1.09,\qquad
\alpha_{\rm worst}\approx 0.96.
\]
The median value $\alpha_{\rm median}\approx 1.31$ is close to the geometric
optimum $4/3$.  This agreement is not itself an arithmetic improvement: for any
well-spread configuration with $|P_k|\asymp 5^{2k}$, the Haar-typical scale
$N^{-2/3}$ translates tautologically into $5^{-4k/3}$.  Thus the median tracks
the geometric baseline rather than beating it. The tail exponents are smaller, and all four exceed the
threshold \(\alpha=2/3\) needed to improve
\(K(T)\le 2\).  Thus the sampled upper tail is \emph{consistent} with the
plausibility of the refined shell-distribution conjecture, but it is not a proof
of the required supremum estimate.  However, the
conjecture is a uniform
supremum statement over all \(u\in S^3\), so the full deterministic question
requires controlling rare holes beyond what finite random sampling can certify.
We develop these numerical diagnostics in detail in Part~II.


\section{Random-target compilation diagnostics for quaternion shells}
\label{sec:exp}
\subsection{Integer quaternions and lattice shells}
Let $\mathbb{H}(\mathbb{Z})$ denote the integer quaternions $q=a+bi+cj+dk$
with $(a,b,c,d)\in\mathbb{Z}^4$. The norm is $|q|^2=a^2+b^2+c^2+d^2$.
For $k\in\mathbb{N}$, define the \emph{shell}
\[
\mathcal{Q}_k := \{(a,b,c,d)\in\mathbb{Z}^4:\ a^2+b^2+c^2+d^2 = 5^{2k}\}.
\]
Normalize to obtain a finite subset of $S^3$:
\begin{equation}\label{eq:Pk}
P_k := \left\{ \frac{1}{5^k}(a,b,c,d)\ :\ (a,b,c,d)\in \mathcal{Q}_k \right\}
\subset S^3.
\end{equation}
In practice, $\mathcal{Q}_k$ is enumerated by precomputing all pairs
$(x,y)\in\mathbb{Z}^2$ with $x^2+y^2=s$ for $s\le 5^{2k}$ and combining
$(a,b)$ with $(c,d)$ so that $(a^2+b^2)+(c^2+d^2)=5^{2k}$.

The shells \(P_k\) are the complete norm-\(5^{2k}\) shells:
\[
|P_k|=2(5^{2k+1}-1),
\]
giving \(248,6248,156248,3906248\) for \(k=1,2,3,4\).  They should not be identified
with a single exact word sphere \(U_T(k)\).  Rather, after projection to
\(PSU(2)\), every point of \(P_k\) has word height at most \(2k\), by Lemma~\ref{lem:shell_lifting}.  The experiments use
\(P_k\) because it is the full shell accessible to enumeration; the probabilistic
analysis of Part~II applies to \(P_k\) directly, and the typical error scale from
Section~\ref{sec:union} provides a baseline against which the covering quality of
\(P_k\) is measured.

\subsection{Empirical Error Distributions on the Shells}
\label{subsec:empirical-shells}

The numerical diagnostic measures the full empirical error distribution
of the quaternion shells.  This distributional view is important because
different statistics probe different mathematical regimes: the median
measures Haar-typical approximation, upper quantiles measure the
beginning of rare-event behavior, and the deterministic covering radius
is governed by the supremum over all targets.

We enumerated the complete quaternion shells
\[
P_k
=
\left\{
\frac{x}{5^k}\in S^3 :
x\in\mathbb Z^4,\;
|x|^2=5^{2k}
\right\}
\]
for \(k=1,2,3,4\), giving
\[
|P_1|=248,\qquad |P_2|=6248,\qquad |P_3|=156248,\qquad |P_4|=3906248.
\]
These are the even norm levels of the integer-quaternion problem.  The
choice \(|x|^2=5^{2k}\) ensures that \(x/5^k\in S^3\).

For each shell we sampled \(m=20\,000\) Haar-random targets
\(u\in S^3\) and computed
\[
S_k(u)
=
\max_{p\in P_k}\langle u,p\rangle,
\qquad
\mathrm{err}_k(u)
=
1-S_k(u).
\]
We use this quantity because the metric on \(SU(2)\cong S^3\) is
directly determined by the inner product.  Earlier we defined
\[
d_G(M,N)
=
\sqrt{
1-\frac{|\mathrm{Tr}(M^\dagger N)|}{2}
},
\]
and under the identification \(SU(2)\cong S^3\), the trace term is the
absolute value of the standard Euclidean inner product:
\[
\frac{|\mathrm{Tr}(M^\dagger N)|}{2}=|\langle u,p\rangle|.
\]
Thus the projective distance satisfies \(d_G(u,p)^2=1-|\langle u,p\rangle|\).
Since the full shell \(P_k\) is centrally symmetric (\(p\in P_k\Rightarrow -p\in
P_k\)), we have
\[
\max_{p\in P_k}|\langle u,p\rangle|
=
\max_{p\in P_k}\langle u,p\rangle.
\]
Therefore minimizing the projective distance between a target \(u\) and the shell
\(P_k\) is equivalent to maximizing the ordinary inner product over \(P_k\), which
is why we define the approximation error by
\[
\mathrm{err}_k(u)
=
1-\max_{p\in P_k}\langle u,p\rangle.
\]

We report the empirical median, the \(99\%\) quantile, the \(99.9\%\)
quantile, and the worst sampled error.  We also define the effective
proxy exponent
\[
K_{\mathrm{proxy}}
=
-\frac{\log(\mathrm{err})}{\log N},
\qquad
N=|P_k|,
\]
so that \(\mathrm{err}\asymp N^{-K_{\mathrm{proxy}}}\). With this normalization the
generic Haar-typical scale \(\mathrm{err}\asymp N^{-2/3}\) corresponds to
\(K_{\mathrm{proxy}}=2/3\); larger errors give smaller exponents.
This exponent is a descriptive statistic for a sampled quantile and is
used to compare different parts of the empirical distribution.

\begin{table}[H]
\centering
\caption{Empirical error statistics for the quaternion shells \(P_k\).
Statistics are computed from \(m=20\,000\) Haar-random targets on \(S^3\).  The
\(k=4\) row uses exact shell enumeration and nearest-neighbor search in
\(\mathbb R^4\).}
\label{tab:numerics}
\small
\begin{tabular}{cccccccc}
\toprule
\(k\) & \(N\) &
median &
\(q_{0.99}\) &
\(q_{0.999}\) &
worst &
median \(K_{\rm proxy}\) &
worst \(K_{\rm proxy}\)
\\
\midrule
1 &
248 &
0.025170 &
0.059945 &
0.076637 &
0.092038 &
0.667843 &
0.432681
\\
2 &
6248 &
0.003195 &
0.012308 &
0.016944 &
0.019948 &
0.657455 &
0.447897
\\
3 &
156248 &
0.000375 &
0.001537 &
0.002332 &
0.004210 &
0.659625 &
0.457413
\\
4 &
3906248 &
0.0000438 &
0.0001679 &
0.0002629 &
0.0007370 &
0.661198 &
0.475222
\\
\bottomrule
\end{tabular}
\end{table}

The table shows three features.

First, the median errors sit close to the generic Haar-typical scale.
The cap/union analysis of Section~\ref{sec:union} shows that no \(N\)-point set
can do appreciably better than \(N^{-2/3}\) for a typical target, while
Theorem~\ref{thm:random-benchmark} shows that independent Haar points have median
\[
        1.101\,N^{-2/3}+o(N^{-2/3}).
\]
The observed scaled medians
\[
        \mathrm{median}(\mathrm{err}_k)\,N^{2/3}
        =0.994,\;1.084,\;1.088,\;1.087\qquad(k=1,2,3,4)
\]
are very close to this random benchmark.  Equivalently, the proxy exponents
\(0.668,\,0.657,\,0.660,\,0.661\) are all close to \(2/3\).  Thus the shells attain the
optimal typical scale: they approximate Haar-typical targets about as well as a
well-distributed random configuration.  This is the correct reading; there is
\emph{no} typical-target gain beyond the geometric optimum, and the energy
diagnostic below gives independent evidence that the shells are globally well spread.

Second, the upper quantiles are separated from the median, recording the
rare-event regime. Under the normalization \(\mathrm{err}\asymp N^{-K_{\rm proxy}}\),
larger errors give \emph{smaller} exponents.  Across \(k=1,2,3,4\), the \(99\%\)
quantiles have \(K_{\rm proxy}\approx 0.50\)–\(0.57\), the \(99.9\%\) quantiles
have \(K_{\rm proxy}\approx 0.47\)–\(0.54\), and the sampled-worst errors have
\(K_{\rm proxy}\approx 0.43\)–\(0.48\). Equivalently, in covering-exponent normalization
(\(\mu(B_G(\varepsilon))\asymp\varepsilon^3\), \(\mathrm{err}\asymp\varepsilon^2\),
so that the relevant ratio is \(\tfrac{2\log N}{3\log(1/\mathrm{err})}\)), the
sampled-worst quantile corresponds to an effective covering exponent
\(\approx 1.40\)–\(1.54\), lying between the proven bounds \(4/3\) and \(2\). This
upper-tail information is essential for relating the experiments to deterministic
covering, which is controlled by extremal rather than typical targets.

Third, the observed profile is compatible with the arithmetic covering
picture.  The known upper bound \(K(T)\le 2\) follows from the LPS/Chiu
spectral covering argument \cite{LPS86,LPS87,Chiu95} (Theorem~\ref{thm:KT2}).
The lower obstruction \(K(T)\ge 4/3\) follows from the arithmetic big-hole
phenomenon, attributed by Sarnak to Harman~\cite{Harman90} and
reproduced in the quaternion setting in Sarnak's letter~\cite{letter}.
The numerical shell profile is consistent with this: the typical target is
covered at the geometric optimum, while the sampled worst case is consistent
with the conjectured true covering exponent near \(4/3\) (though, being a
quantile, it cannot certify the deterministic supremum).

This finite-shell viewpoint is closely aligned with the local-statistics
literature on lattice points on spheres \cite{BSR12,BRS16}, but focuses on
the gate-approximation functional rather than pair counts, cap variance, or
nearest-neighbor spacing.

Thus the numerical contribution is a typical-to-tail profile for
arithmetic quaternion shells:
\[
\begin{aligned}
&\text{optimal Haar-typical scale }(N^{-2/3})\\
&\qquad \longrightarrow\ \text{rare-event upper tail}
\ \longrightarrow\ \text{deterministic covering obstruction}.
\end{aligned}
\]
This profile motivates Conjecture~\ref{conj:SD}: an improvement from
\(K(T)\le 2\) to \(K(T)\le 2-\delta\) would require arithmetic input
capable of controlling the upper tail, not only the median behavior.

\subsection{Electrostatic Energy Diagnostic}
\label{subsec:energy-diagnostic}

The nearest-neighbor functional \(\mathrm{err}_k(u)\) measures covering from
the point of view of a target \(u\).  As a complementary global statistic, we
also compute the electrostatic Riesz energy
\begin{equation}\label{eq:energy-def}
E(P_k)
=
\sum_{\substack{p,q\in P_k\\p\ne q}}
\frac{1}{\|p-q\|},
\qquad N=|P_k|.
\end{equation}
This statistic is standard in the study of well-distributed points on spheres:
random points, Fekete-type configurations, and minimal-energy configurations
have the same leading-order energy constant for the kernel \(\|x-y\|^{-1}\).
Thus \(E(P_k)\) gives a second test of whether the arithmetic quaternion
shells behave like optimized or random point sets to leading order.

\begin{proposition}[Haar leading constant on \(S^3\)]
\label{prop:haar-energy-constant}
Let \(X,Y\) be independent Haar-random points on \(S^3\subset\mathbb R^4\).
Then
\begin{equation}\label{eq:haar-energy-constant}
\mathbb E\frac{1}{\|X-Y\|}
=
\frac{8}{3\pi}.
\end{equation}
Consequently, for \(N\) independent Haar-random points
\(X_1,\ldots,X_N\in S^3\),
\[
\mathbb E\left[
\frac{1}{N(N-1)}
\sum_{i\ne j}\frac{1}{\|X_i-X_j\|}
\right]
=
\frac{8}{3\pi}.
\]
\end{proposition}

\begin{proof}
By rotational invariance, fix \(Y=e_1\) and write
\(T=\langle X,e_1\rangle\).  For \(X\sim\Haar(S^3)\), the one-dimensional
marginal has density
\[
f_T(t)=\frac{2}{\pi}\sqrt{1-t^2},
\qquad -1\le t\le 1.
\]
Since \(\|X-e_1\|^2=2-2T\), we obtain
\begin{align*}
\mathbb E\frac{1}{\|X-Y\|}
&=
\int_{-1}^{1}
\frac{1}{\sqrt{2-2t}}\,
\frac{2}{\pi}\sqrt{1-t^2}\,dt \\
&=
\frac{2}{\pi}
\int_{-1}^{1}
\sqrt{\frac{(1-t)(1+t)}{2(1-t)}}\,dt  \\
&=
\frac{2}{\pi\sqrt 2}
\int_{-1}^{1}\sqrt{1+t}\,dt.
\end{align*}
With \(s=1+t\), this becomes
\[
\frac{2}{\pi\sqrt 2}
\int_0^2 s^{1/2}\,ds
=
\frac{2}{\pi\sqrt 2}\cdot \frac{2}{3}2^{3/2}
=
\frac{8}{3\pi}.
\]
The second claim follows by linearity of expectation over the ordered pairs
\((i,j)\), \(i\ne j\).
\end{proof}

We computed \(E(P_k)/(N(N-1))\) for \(k=1,2,3\).  For \(k=1,2\), the energy was
computed exactly by summing all ordered distinct pairs.  For \(k=3\), the shell
has \(156248\) points, so we estimated the normalized energy by Monte Carlo
sampling of \(2,000,000\) ordered distinct pairs.  The results are shown in
Table~\ref{tab:energy}.

\begin{table}[H]
\centering
\caption{Electrostatic energy of the quaternion shells.  The normalized energy
\(E(P_k)/(N(N-1))\) is compared with the Haar-random leading constant
\(8/(3\pi)\approx 0.848826\).}
\label{tab:energy}
\small
\begin{tabular}{ccccc}
\toprule
\(k\) & \(N=|P_k|\) &
\(E(P_k)/(N(N-1))\) &
\(8/(3\pi)\) &
relative error \\
\midrule
1 & 248    & 0.829913 & 0.848826 & \(-2.23\%\) \\
2 & 6248   & 0.847577 & 0.848826 & \(-0.15\%\) \\
3 & 156248 & 0.848877 & 0.848826 & \(0.006\%\) \\
\bottomrule
\end{tabular}
\end{table}

These computations are numerically consistent with the asymptotic energy relation
\begin{equation}\label{eq:shell-energy-law}
E(P_k)
=
\frac{8}{3\pi}|P_k|^2
+
o\!\left(|P_k|^2\right),
\end{equation}
which would assert, after dividing by \(|P_k|^2\), that
\(E(P_k)/|P_k|^2\to 8/(3\pi)\) as \(k\to\infty\). We emphasize that
\eqref{eq:shell-energy-law} is a conjectural reading of three data points, not a
theorem: establishing it would require quantitative equidistribution of the
shells strong enough to control the energy functional, which we do not prove
here. Subject to that caveat, the shells appear to share the leading-order energy
statistic of Haar-random points on \(S^3\).

This should be interpreted in the spirit of the local-statistics program of
Bourgain--Sarnak--Rudnick \cite{BSR12,BRS16}.  Their viewpoint is that arithmetic lattice points
on spheres should be compared against random point processes using statistics
such as electrostatic energy, nearest-neighbor spacing, Ripley statistics,
covering radius, and cap variance.  The question is not whether the arithmetic
points are literally random, they are highly structured, but whether their
large-scale statistics behave like those of an optimally distributed point set.

The quaternion shells provide a striking example of this phenomenon.  The points
are not produced by minimizing an energy functional or solving a geometric
optimization problem.  Instead, they are forced by the arithmetic constraint
\[
a^2+b^2+c^2+d^2 = 5^{2k}.
\]
Nevertheless, the normalized shell energy converges numerically to the same
constant \(8/(3\pi)\) that governs Haar-random configurations.  In this sense,
the shells behave globally like highly optimized or random point sets on
\(S^3\).

At the same time, the covering-error quantiles in
Table~\ref{tab:numerics} show that rare extremal holes still persist.  The
median sits at the geometric optimum \(\asymp N^{-2/3}\), while the upper tail
remains separated from it.
This is exactly the tension emphasized in the Bourgain--Sarnak--Rudnick
philosophy: arithmetic point sets can exhibit consistent with the random benchmark leading-order
statistics globally, while still possessing rare deterministic obstructions at
extreme scales.  In the present setting, the shells appear globally optimized
from the viewpoint of energy, and the median covering error attains the generic
Haar-typical scale, yet the extremal covering behavior is still
governed by rare large holes.

\subsection{Concentration of the maximum correlation}\label{ssc:concentration}

\begin{lemma}[L\'{e}vy concentration for maximum correlations]\label{lem:levy}
Let $P\subset S^3$ be finite and define
\[
        f(u)=\max_{p\in P}\langle u,p\rangle,
        \qquad u\in S^3.
\]
Then $f$ is $1$-Lipschitz with respect to the Euclidean metric on $S^3$.
Consequently, there are absolute constants $c,C>0$ such that, for every
$t>0$,
\[
        \Haar\bigl(|f(u)-\mathbb E f(u)|\ge t\bigr)
        \le C\exp(-c t^2).
\]
\end{lemma}

\begin{proof}
For $u,v\in S^3$,
\[
\begin{aligned}
|f(u)-f(v)|
&=\left|\max_{p\in P}\langle u,p\rangle
      -\max_{p\in P}\langle v,p\rangle\right|  \\
&\le \max_{p\in P}|\langle u-v,p\rangle|
 \le \|u-v\|_2,
\end{aligned}
\]
since each $p\in P$ has unit norm.  Thus $f$ is $1$-Lipschitz.  The stated
probability bound is the standard L\'{e}vy concentration inequality for
Lipschitz functions on the sphere, with the dimension fixed at $4$; see, for
example, \cite[Chapter~5]{VershyninHDP}.
\end{proof}

\paragraph{Interpretation of concentration.}
Lemma~\ref{lem:levy} does not determine the approximation scale of the maximum
correlation; rather, it controls \emph{fluctuations} around that scale. Since
\[
f(u)=\max_{p\in P}\langle u,p\rangle
\]
is $1$-Lipschitz on $S^3$, L\'{e}vy concentration implies that once the mean
$\mathbb{E}f(u)$ satisfies $\mathbb{E}f(u)\approx 1-\eta$, deviations of order
$t$ away from $\mathbb{E}f(u)$ occur with probability at most $\exp(-c\,t^2)$.

Consequently, the mean, median, and other central quantiles of $f(u)$ coincide
up to constants, and $f(u)$ is tightly concentrated for Haar-random $u$.
This explains why Monte-Carlo histograms of the best inner product are narrow
and stable, and why empirical ``typical'' performance can be read off from the
mean. However, concentration alone does not identify the value of $\eta$;
that scale is set by spherical cap probabilities and a union bound, which we
turn to next.

\begin{figure}[H]
\centering
\begin{subfigure}[t]{0.49\linewidth}
  \centering
  \safeincludegraphics[width=\linewidth]{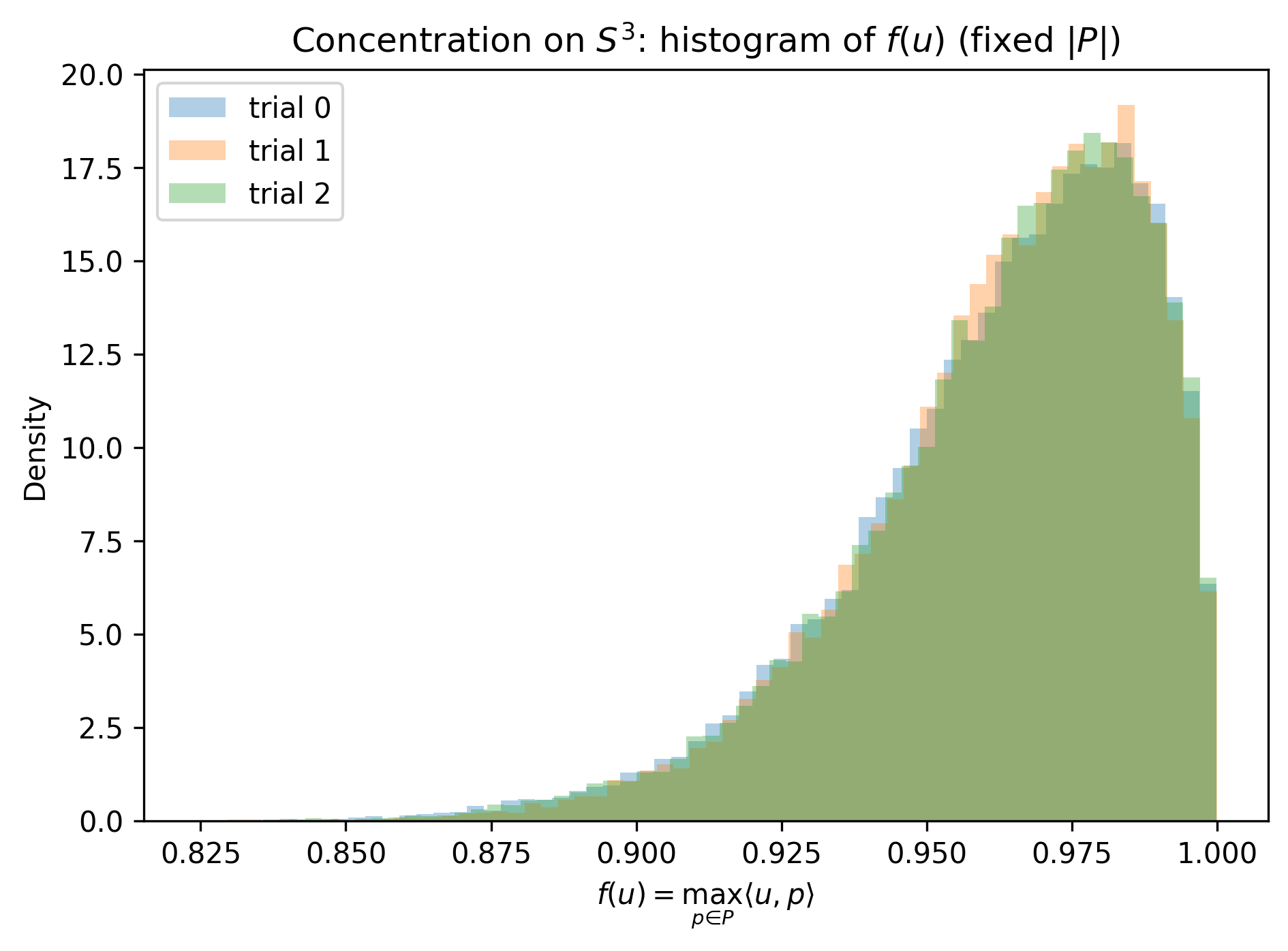}
  \caption{Histogram of $f(u)$ across independent trials.}
  \label{fig:conc_hist}
\end{subfigure}\hfill
\begin{subfigure}[t]{0.49\linewidth}
  \centering
  \safeincludegraphics[width=\linewidth]{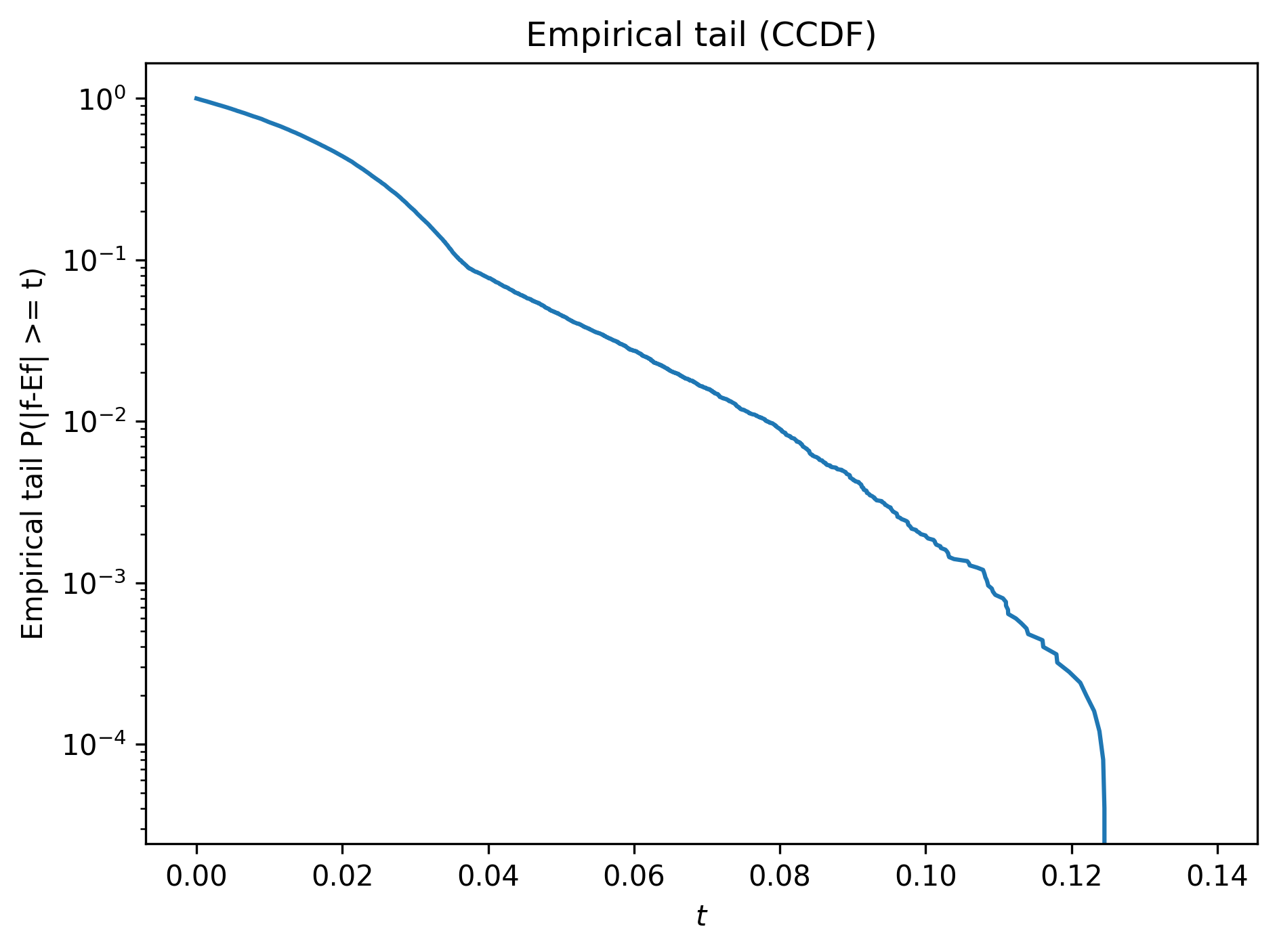}
  \caption{Empirical two-sided tail $\Pr(|f(u)-\mathbb E f(u)|\ge t)$.}
  \label{fig:conc_tail}
\end{subfigure}

\vspace{0.5em}

\begin{subfigure}[t]{0.49\linewidth}
  \centering
  \safeincludegraphics[width=\linewidth]{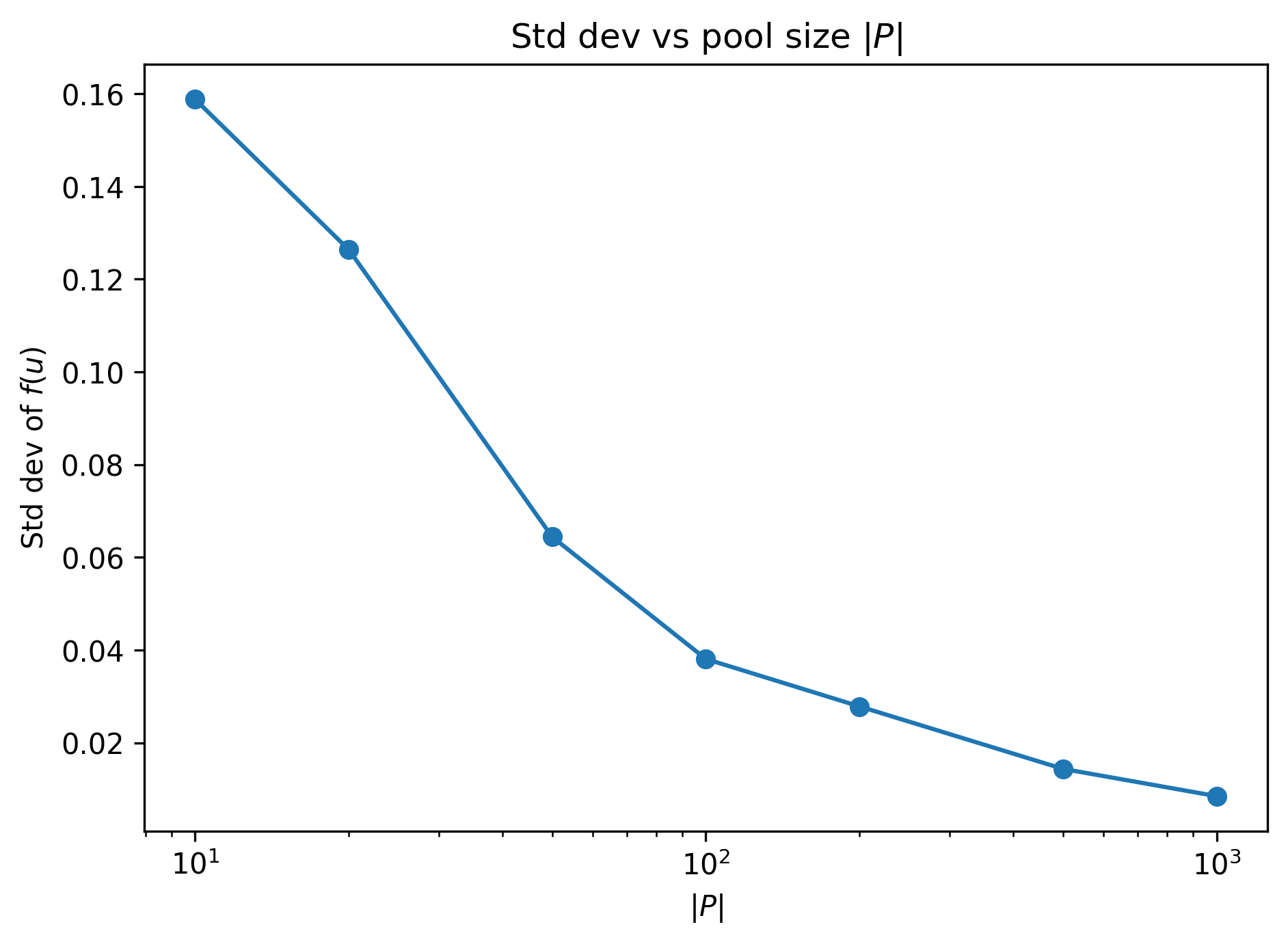}
  \caption{Standard deviation of $f(u)$ vs.\ pool size $|P|$.}
  \label{fig:conc_std}
\end{subfigure}\hfill
\begin{subfigure}[t]{0.49\linewidth}
  \centering
  \safeincludegraphics[width=\linewidth]{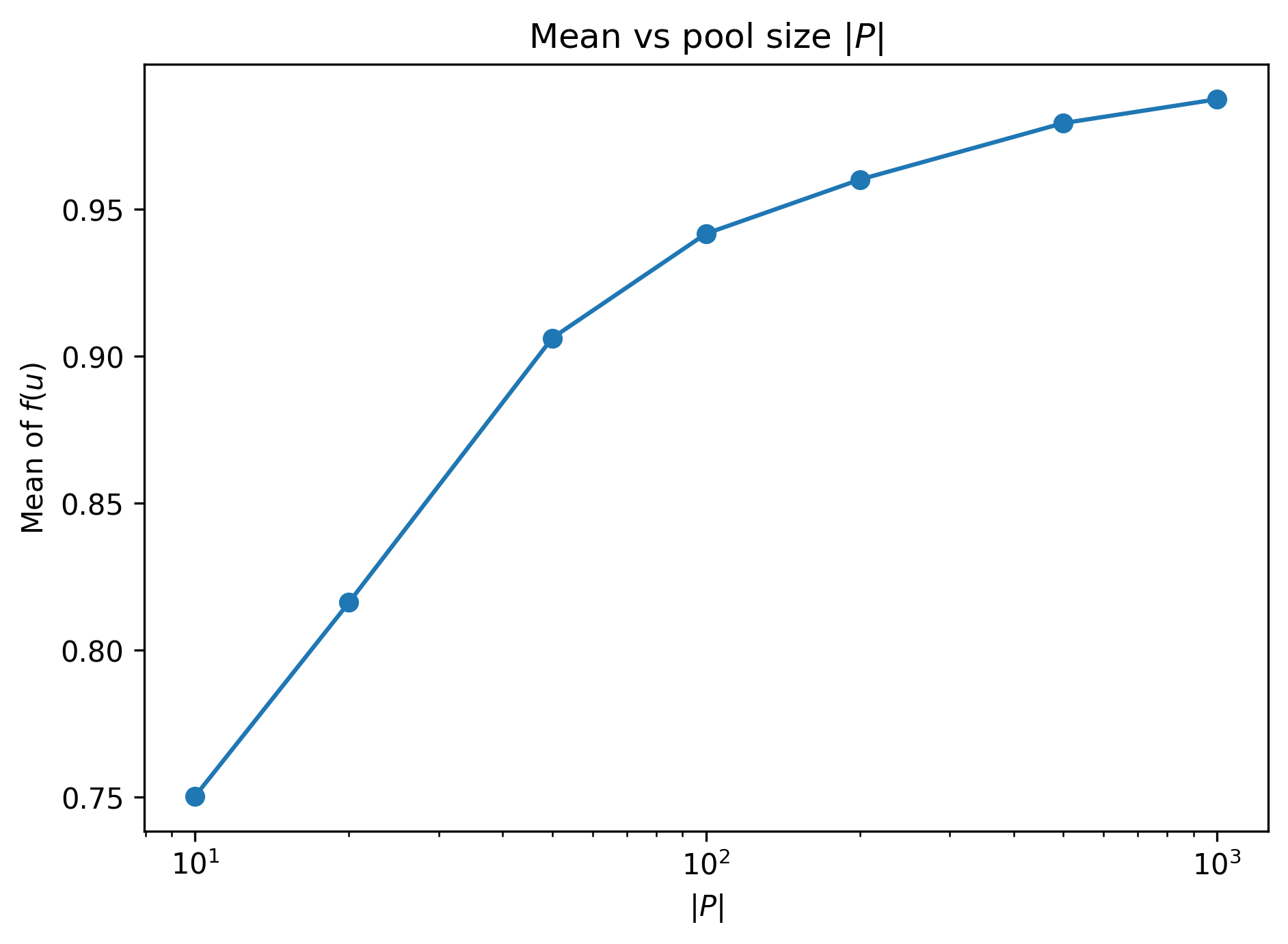}
  \caption{Mean of $f(u)$ vs.\ pool size $|P|$.}
  \label{fig:conc_mean}
\end{subfigure}

\caption{\textbf{Concentration of the maximum correlation on $S^3$.}
Let $u\sim\Haar$ on $S^3\subset\mathbb R^4$ and fix a finite set $P\subset S^3$. We
study $f(u)=\max_{p\in P}\langle u,p\rangle$, which is $1$-Lipschitz as a
maximum of $1$-Lipschitz linear functionals. Panels (a)--(b) show that $f(u)$
is tightly concentrated around its typical value (narrow histogram and rapidly
decaying empirical tails). Panels (c)--(d) separate \emph{scale} from
\emph{fluctuations}: increasing $|P|$ raises the typical value $\mathbb E f(u)$ (d),
while the fluctuations around that value remain small (c), consistent with
L\'{e}vy concentration on the sphere.}
\label{fig:concentration_S3}
\end{figure}

\section{Spherical Cap Probabilities on \texorpdfstring{$S^3$}{S3}}\label{sec:haar}
The probabilistic baseline in Part~II uses only the elementary Haar measure of a
small cap on \(S^3\).  We record the calculation in the normalization used for
our trace-defect error.

Fix \(p\in S^3\subset\mathbb{R}^4\) and let \(u\sim \Haar\) be uniform on
\(S^3\).  Define \(X:=\ip{u}{p}\).  By rotational invariance we may assume
\(p=e_1\), hence \(X=u_1\).  For \(u\sim\mathrm{Unif}(S^{n-1})\), a
one-dimensional marginal has density proportional to \((1-t^2)^{(n-3)/2}\) on
\([-1,1]\) \cite[Exercise~3.27(b)]{VershyninHDP}.  For \(n=4\), normalization by
\(\int_{-1}^1\sqrt{1-t^2}\,dt=\pi/2\) gives
\begin{equation}\label{eq:densityS3}
f_X(t)=\frac{2}{\pi}\sqrt{1-t^2},\qquad t\in[-1,1].
\end{equation}
Consequently, for \(\eta\in(0,1)\),
\begin{equation}\label{eq:cap_bound}
\Haar\bigl(\ip{u}{p}\ge 1-\eta\bigr)
=
\int_{1-\eta}^1 \frac{2}{\pi}\sqrt{1-t^2}\,dt
\le C_3\eta^{3/2},
\qquad C_3=\frac{4\sqrt2}{3\pi}.
\end{equation}
Indeed, for \(t\in[1-\eta,1]\),
\(1-t^2=(1-t)(1+t)\le2(1-t)\), and integration gives the stated constant.  The
same expansion also gives the sharp small-cap asymptotic
\begin{equation}\label{eq:cap_asymp}
\Haar\bigl(\ip{u}{p}\ge 1-\eta\bigr)
= C_3\eta^{3/2}+O(\eta^{5/2}),
\qquad C_3=\frac{4\sqrt2}{3\pi}.
\end{equation}
This cap estimate, rather than a general concentration inequality, is the input
used in the union-bound and random-benchmark results below.

\section{A Union-Bound Control and the Haar-Typical Error Scale}\label{sec:union}

The cap calculation in Section~\ref{sec:haar} identifies the scale of a typical
nearest-neighbor error.  For
\[
S_P(u):=\max_{p\in P}\langle u,p\rangle,
\]
the event \(S_P(u)\ge1-\eta\) occurs only if the Haar-random target falls in one
of the \(N=|P|\) caps of height \(\eta\) centered at the points of \(P\).  A
union bound therefore gives the universal Haar-typical scale, independent of any
arithmetic structure of \(P\).

\begin{lemma}[Union bound via caps]\label{lem:union}
Let $P\subset S^3$ with $|P|=N$, and let $u\sim\Haar$ be uniform on $S^3$.
Then for any $\eta\in(0,1)$,
\begin{equation}\label{eq:union_bound}
\Haar\Big(S_P(u)\ge 1-\eta\Big)
=
\Haar\Big(\max_{p\in P}\ip{u}{p}\ge 1-\eta\Big)
\le N\,\Haar\big(\ip{u}{p_0}\ge 1-\eta\big)
\le C_3\,N\,\eta^{3/2},
\end{equation}
where $p_0\in S^3$ is any fixed unit vector and $C_3=\frac{4\sqrt{2}}{3\pi}$
is the constant from \eqref{eq:cap_bound}.
\end{lemma}

\begin{proof}
By the union bound,
\[
\Haar\Big(\exists\,p\in P:\ \ip{u}{p}\ge 1-\eta\Big)
\le \sum_{p\in P}\Haar\big(\ip{u}{p}\ge 1-\eta\big).
\]
By rotational invariance of $\Haar$ on $S^3$, the probability
$\Haar(\ip{u}{p}\ge 1-\eta)$ is the same for every fixed $p\in S^3$; hence for
any fixed $p_0\in S^3$ the sum equals
\[
\sum_{p\in P}\Haar\big(\ip{u}{p}\ge 1-\eta\big)
= N\,\Haar\big(\ip{u}{p_0}\ge 1-\eta\big).
\]
Finally, the spherical cap bound \eqref{eq:cap_bound} yields
$\Haar(\ip{u}{p_0}\ge 1-\eta)\le C_3\,\eta^{3/2}$, giving
\eqref{eq:union_bound}.
\end{proof}

\begin{theorem}[Universal Haar-typical lower scale]\label{thm:typical-floor}
Let $P\subset S^3$ have $|P|=N$, let $u\sim\Haar$, and define
\[
        e_P(u)=1-\max_{p\in P}\langle u,p\rangle .
\]
For every $q\in(0,1)$, the lower $q$-quantile
\[
        Q_q(P):=\inf\{\eta>0:\ \Haar(e_P(u)\le\eta)\ge q\}
\]
satisfies
\[
        Q_q(P)\ge \left(\frac{q}{C_3N}\right)^{2/3}.
\]
In particular, every median of $e_P$ is bounded below by
\[
        \operatorname{med}(e_P)\ge \left(\frac{1}{2C_3N}\right)^{2/3}.
\]
Thus no $N$-point set on $S^3$ can have Haar-typical trace-defect error smaller
than a constant multiple of $N^{-2/3}$.
\end{theorem}

\begin{proof}
By Lemma~\ref{lem:union},
\[
        \Haar(e_P(u)\le\eta)
        =\Haar(S_P(u)\ge1-\eta)
        \le C_3N\eta^{3/2}.
\]
If $\eta<(q/(C_3N))^{2/3}$, then the right-hand side is $<q$, so the probability
of error at most $\eta$ is still below $q$.  Hence the $q$-quantile cannot occur
below $(q/(C_3N))^{2/3}$.  Taking $q=1/2$ gives the median bound.
\end{proof}

\begin{theorem}[Independent Haar benchmark]\label{thm:random-benchmark}
Let $P_N=\{p_1,\ldots,p_N\}$ consist of $N$ independent Haar-random points on
$S^3$, independent of $u\sim\Haar$, and define
\[
        e_N=1-\max_{1\le i\le N}\langle u,p_i\rangle .
\]
Then for every fixed $x\ge0$,
\[
        \lim_{N\to\infty}\mathbb P\bigl(N^{2/3}e_N>x\bigr)
        =\exp(-C_3x^{3/2}),
        \qquad C_3=\frac{4\sqrt2}{3\pi}.
\]
Consequently the random-set median satisfies
\[
        \operatorname{med}(e_N)
        \sim \left(\frac{\log 2}{C_3}\right)^{2/3}N^{-2/3}
        \approx 1.101\,N^{-2/3}.
\]
\end{theorem}

\begin{proof}
By rotational invariance, conditional on $u$, the random variables
$\langle u,p_i\rangle$ are independent with the same distribution as
$\langle e_1,p_i\rangle$.  Let
\[
        a(\eta)=\Haar(\langle e_1,p\rangle\ge1-\eta).
\]
Then
\[
        \mathbb P(e_N>\eta)=(1-a(\eta))^N.
\]
By the cap asymptotic~\eqref{eq:cap_asymp},
$a(\eta)=C_3\eta^{3/2}+O(\eta^{5/2})$.  Setting
$\eta=xN^{-2/3}$ gives
\[
        N a(xN^{-2/3})\to C_3x^{3/2}.
\]
Therefore
\[
        (1-a(xN^{-2/3}))^N\to \exp(-C_3x^{3/2}),
\]
which proves the limiting law.  The median follows by solving
$\exp(-C_3x^{3/2})=1/2$.
\end{proof}

\paragraph{Haar-typical scale.}
Lemma~\ref{lem:union} gives $\Haar(S_P(u)\ge 1-\eta)\le C_3 N\eta^{3/2}$, which
is non-negligible only when $\eta\gtrsim N^{-2/3}$. Read contrapositively: if
$\eta\ll N^{-2/3}$, then the probability that \emph{any} point of $P$ lies in a
cap of defect $\eta$ is negligible, so with high probability no $p\in P$ has
$\ip{u}{p}\ge 1-\eta$. In other words, \emph{errors much smaller than
$N^{-2/3}$ occur with probability tending to zero, uniformly for every
$N$-point set}. Quantitatively, the probability is $O(N\eta^{3/2})$, which tends
to zero when $\eta\ll N^{-2/3}$. This makes $N^{-2/3}$ a \emph{lower} limit on
the achievable typical error: the scale $\eta\asymp N^{-2/3}$ is where positive
probability of small error first becomes possible.

The matching \emph{upper} bound -- that typical error is at most $O(N^{-2/3})$ --
requires a lower bound on cap-hitting probability.  Theorem~\ref{thm:random-benchmark}
proves this sharply for independent Haar points and gives the benchmark median
constant $1.101$.  For clustered $P$, the typical error can be much larger than
$N^{-2/3}$.  For the arithmetic shells $P_k$, the experiments show that the
median sits at $\eta_{\mathrm{typ}}\asymp N^{-2/3}$ and even matches the random
benchmark constant closely.

Combining both directions, for a well-spread $N$-point set the Haar-typical
error satisfies
\begin{equation}\label{eq:eta_scale}
\eta_{\mathrm{typ}} \asymp N^{-2/3},
\end{equation}
and in angular distance $\theta(u,P)$ (with $1-\max_{p}\ip{u}{p}\asymp\theta(u,P)^2$
for small $\theta$), this corresponds to $\theta_{\mathrm{typ}}(u,P)\asymp N^{-1/3}$.
The bound on the small-error probability from Lemma~\ref{lem:union} is a
constraint on \emph{how small} the typical error can be; it does not bound
how large it is.

\begin{remark}
Throughout, $\mu(B_G(\varepsilon))\asymp\varepsilon^3$ so a ball of volume
$N^{-a}$ has metric radius $\varepsilon\asymp N^{-a/3}$ and trace-defect
$1-\cos\theta\asymp\varepsilon^2\asymp N^{-2a/3}$. We summarize the scales:
\begin{itemize}[leftmargin=2em,itemsep=2pt]
\item \emph{Generic Haar-typical floor} (Lemma~\ref{lem:union}): for any
  $N$-point set, errors $\eta\ll N^{-2/3}$ have negligible probability for a
  Haar-random target; thus $N^{-2/3}$ is a lower limit on the achievable
  typical error. Well-spread sets also attain the matching upper bound, so
  $\eta_{\mathrm{typ}}\asymp N^{-2/3}$ (metric radius $\asymp N^{-1/3}$) is
  the \emph{optimal} generic typical scale, not merely an upper bound.
\item \emph{Proven deterministic covering} (Theorem~\ref{thm:KT2},
  $K(T)\le 2$): every target covered at volume scale $N^{-1/2}$, i.e.\ metric
  radius $N^{-1/6}$ and trace-defect $N^{-1/3}$ (using $a=1/2$).
  Since $N^{-1/6}>N^{-1/3}$, this guarantee is coarser in radius than the
  typical scale; it ensures all targets are covered but at a larger error.
\item \emph{Conjectured true covering} \cite{BKS16}: empty balls of volume
  $N^{-3/4}$ are conjectured to exist, giving covering radius
  $\varepsilon\asymp N^{-1/4}$ and trace-defect $\asymp N^{-1/2}$ (using $a=3/4$).
  This is coarser than the typical scale but finer than the proven $K(T)\le 2$
  guarantee; it corresponds to the Hecke equidistribution conjecture
  $K(T)=4/3$.
\item \emph{Numerical worst-case quantile}: the sampled worst error in
  Table~\ref{tab:numerics} has trace-defect $\asymp N^{-1/2}$, i.e.\ metric
  radius $\asymp N^{-1/4}$, consistent with the conjectured scale. This is an
  upper quantile of the error distribution, not the deterministic covering
  radius.
\end{itemize}
Smaller metric radius (larger exponent in \(N^{-(\cdot)}\)) is better.  Ordered
from finest to coarsest radius, the hierarchy is: typical scale \(N^{-1/3}\),
conjectured worst case \(N^{-1/4}\), and proven worst-case guarantee
\(N^{-1/6}\).
\end{remark}

\subsection{Specialization to lattice shells \texorpdfstring{$P=P_k$}{P=Pk}}
For $P=P_k$ with $|P_k|=N_k$, the floor/optimality analysis of \eqref{eq:eta_scale} gives
\begin{equation}\label{eq:typ_k}
\eta_{\mathrm{typ}}(P_k) \asymp N_k^{-2/3}
\qquad\text{(Haar-typical scale, valid for well-spread $P_k$)}.
\end{equation}
The shell medians (Table~\ref{tab:numerics}) confirm that $P_k$ is well-spread:
errors $\eta\ll N_k^{-2/3}$ are suppressed and the median sits at
$\asymp N_k^{-2/3}$. There is no typical-target gain beyond the geometric optimum.
If heuristically $N_k \asymp 5^{\beta k}$ for some $\beta>0$, the typical scale becomes
\begin{equation}\label{eq:typ_exp}
\eta_{\mathrm{typ}}(P_k) \asymp 5^{-(2\beta/3)k}.
\end{equation}

\subsection{What Monte--Carlo ``worst case'' does (and does not) certify}

Let $u_1,\dots,u_m\stackrel{iid}{\sim}\Haar$ and define the sample worst error
\[
\widehat{\rho}_m(P):=\max_{1\le i\le m}\mathrm{err}_P(u_i).
\]
If $F(\eta)=\Haar(\mathrm{err}_P(u)\le \eta)$ denotes the Haar CDF of the
error, then
\[
\Haar\!\left(\widehat{\rho}_m(P)\le \eta\right)=F(\eta)^m.
\]
Thus $\widehat{\rho}_m(P)$ estimates an \emph{upper quantile} of
$\mathrm{err}_P(u)$, approximately the $(1-1/m)$-quantile, rather than the
deterministic covering radius $\rho(P)=\sup_{u\in S^3}\mathrm{err}_P(u)$.  This average-versus-worst-case distinction is the only point needed here: random
sampling can estimate high Haar quantiles, but it does not certify an extremal
direction.

\paragraph{Numerical interpretation.}
In our experiments (Table~\ref{tab:numerics}), the observed sample worst errors
\[
\mathrm{worst\_err}_1 \approx 9.2\times 10^{-2},\quad
\mathrm{worst\_err}_2 \approx 2.0\times 10^{-2},\quad
\mathrm{worst\_err}_3 \approx 4.2\times 10^{-3}
\]
correspond to the largest errors among $m$ Haar-random targets.
For example, when $k=3$ and $|P_3|\approx 1.56\times 10^5$, the value
$\widehat{\rho}_m(P_3)\approx 4.2\times 10^{-3}$ should be interpreted as the
error level exceeded by only a fraction $\approx 1/m$ of Haar-random targets,
not as a certified upper bound on $\sup_u \mathrm{err}_{P_3}(u)$.

Equivalently, if $\widehat{\rho}_m(P_3)=\eta$, then
\[
F(\eta)\approx 1-\frac{1}{m},
\]
so increasing $m$ tightens the estimated \emph{quantile} but does not rule out
the existence of rare targets $u$ (of Haar measure $\ll 1/m$) with
significantly larger error.

\subsection{Comparison with the deterministic shell target}
Conjecture~\ref{conj:SD} is a deterministic shell-covering statement in
trace-defect scale.  In the notation of Section~\ref{ssc:RUB}, it asserts that
there exist constants \(\alpha>2/3\) and \(C>0\) such that
\[
\forall\,u\in S^3\quad \exists\,p\in P_k:
\qquad
1-\ip{u}{p}\le C5^{-\alpha k}.
\]
Equivalently,
\[
\rho(P_k)
:=
\sup_{u\in S^3}
\left(1-\max_{p\in P_k}\ip{u}{p}\right)
\le C5^{-\alpha k}.
\]
This is strictly stronger than the Haar-typical estimate
\eqref{eq:typ_exp}: it controls the supremum over all targets \(u\), whereas
\eqref{eq:typ_exp} and Lemma~\ref{lem:union} only control typical or quantile
behavior under Haar-random sampling.

The threshold \(\alpha>2/3\) is exactly the threshold needed to improve the
known upper bound \(K(T)\le 2\), since the argument in
Section~\ref{ssc:RUB} gives
\[
K(T)\le \frac{4}{3\alpha}.
\]
Thus any deterministic shell estimate with \(\alpha>2/3\) would imply
\(K(T)<2\).  The numerical upper-tail statistics support this conjectural
picture, but do not prove it: they show decay of high sampled quantiles, while
Conjecture~\ref{conj:SD} requires uniform control of the worst uncovered
regions.  Establishing such a bound would require additional arithmetic
equidistribution input beyond the cap-volume baseline, in the spirit of the
Ramanujan-type spectral and strong-approximation methods used in
\cite{OS,EP22,DEP25}.

\section{Conclusion}\label{sec:conclusion}

This paper studies the $p=5$ quaternionic, equivalently Clifford$+V$, gate set as an arithmetic model for single-qubit compilation.  The main message is that random-target approximation and deterministic worst-case synthesis are controlled by different phenomena.  The complete quaternion shells behave almost like well-distributed random point sets at the median, but the worst-case covering exponent is governed by rare arithmetic holes.

The deterministic result explains why the known exponent range
\[
        \frac43\le K(T)\le2
\]
has been difficult to improve.  The positive-kernel barrier shows that Deligne--LPS square-root spectral control, when combined only with a positive cap kernel, certifies covering only at the volume-squared threshold
\[
        |V_T(t)|\gg \mu(B_G(\varepsilon))^{-2}.
\]
This is exactly exponent $2$.  Hence a proof of $K(T)<2$ for worst-case Clifford$+V$ compilation cannot be obtained by sharpening the same diagonal positivity argument; it must use cancellation in localized off-diagonal arithmetic counts.  The twisted-Linnik framework of Browning--Kumaraswamy--Steiner supplies such an input conditionally and gives the endpoint $K(T)=4/3$, matching Harman's obstruction.

The shell-to-gate conversion turns this arithmetic obstruction into a concrete covering target.  A deterministic estimate
\[
        \rho(P_k)\le C5^{-\alpha k}
\]
would imply
\[
        K(T)\le \frac{4}{3\alpha}.
\]
Thus $\alpha>2/3$ is exactly the threshold needed to beat the unconditional exponent $2$, while $\alpha=1$ corresponds to the conditional endpoint.  This gives a precise benchmark for future number-theoretic work on arithmetic single-qubit gate synthesis.

The numerical experiments give a complementary quantum-compilation perspective.  For $P_1,P_2,P_3,P_4$, the median Haar-random target error follows the optimal $N^{-2/3}$ trace-defect scale predicted for well-distributed points on $S^3$.  This indicates that quaternionic gate sets are already strong for typical random one-qubit targets.  The upper tail, however, remains visibly separated from the median, which is consistent with the arithmetic-hole mechanism responsible for worst-case lower bounds.  In short, average-looking performance is not evidence of a worst-case covering theorem.

The open problem left by this work is to prove a uniform rare-hole estimate for the complete quaternion shells, or equivalently prove enough localized off-diagonal cancellation to control $\sup_u\operatorname{err}_{P_k}(u)$.  Such an estimate would directly improve worst-case Clifford$+V$ compilation bounds.  Conversely, the barrier theorem explains why purely spectral equidistribution, without this additional local arithmetic input, is insufficient.

\section*{Data and code availability}

The numerical data reported in this manuscript are generated from exact enumeration of the integer solutions defining the shells $P_k$, followed by nearest-neighbor evaluation on Haar-random targets.  No proprietary or experimental data are used.  The enumeration procedure, error metrics, and random-target interpretation are described in Sections~\ref{sec:exp}--\ref{sec:union}; the authors are happy to provide scripts and raw tables upon request.

\end{document}